\documentclass[prb,showpacs,amsmath,amssymb,twoside,twocolumn]{revtex4}
\usepackage{epsfig}
\usepackage{wasysym}
\usepackage{ae}
\usepackage[T1]{fontenc}

\newcommand{\dd}{\mathrm{d}}
\newcommand{\sign}{\mathrm{sign}}
\renewcommand{\vr}{\mathbf{r}}

\begin{document}

\title{Forces between chemically structured substrates mediated by critical fluids}
\date{\today}
\author{Monika Sprenger}
\author{Frank Schlesener}
\author{S. Dietrich}
\affiliation{Max-Planck-Institut f\"ur Metallforschung, Heisenbergstr. 3, D-70569 Stuttgart, Germany, }
\affiliation{Institut f\"ur Theoretische und Angewandte Physik, Universit\"at Stuttgart, 
             Pfaffenwaldring 57, D-70569 Stuttgart, Germany}

\begin{abstract}
  We consider binary liquid mixtures close to their critical points confined by two parallel,
  geometrically flat but chemically structured, substrates. Universal order parameters profiles are 
  calculated within mean field theory for periodic patterns of stripes with alternating preferences 
  for the two species of the mixture and with different relative positions of the two substrates. 
  From the order parameters profiles the effective forces between the two plates are derived. The 
  tuning of Casimir amplitudes is discussed.
\end{abstract}

\maketitle

\section{\label{sec:intro}Introduction}
Since chemical structuring of surfaces in the $\mu$m-range and below have become feasible, e.g., via 
self-assembled monolayers (SAM) of block-copolymers or mixtures of polymers,\mbox{\cite{Bates1991, 
Krausch2002, Boltau1998}} micro-contact printing,\cite{Kumar1993, Xia1999} or exposure through a mask 
of photosensitive surfaces or surfaces which change their properties after irradiation,\cite{Zhao2002, 
Wang1997, Wang1998, Seki2004} chemically patterned substrates have gained both practical and theoretical 
interest. Chemically structured substrates have applications in micro-reactors, the laboratory on the 
chip, and in chemical sensors.\cite{Thorsen2002, Juncker2002} For these systems the local properties 
of fluids as confined by structured walls play an important role. In turn, wetting \cite{Bauer1999, 
Bauer1999a,Brinkmann2002, Schneemilch2003} as well as critical adsorption of fluids \cite{Sprenger2005} 
at chemically structured substrates are in the focus of theoretical studies, the latter being the basis 
for the present analysis.
While the above studies deal with a bulk phase in contact with a single substrate one is also 
interested in thin films confined between two substrates. Examples are liquids confined in (slit) 
pores or thin -- albeit still large on the microscopic scale -- wetting films for which the vapor phase 
plays the role of the second substrate. For instance there are studies of confined liquid 
crystals,\cite{Kondrat2005} confined block copolymers,\cite{Wang2005, Tsori2001, Tsori2001a, Tsori2003} 
and spherical particles with attractive interactions confined between two structured 
substrates.\cite{Diestler1994, Bock2000, Bock2001, Bock2001a,Greberg2001, Overduin2002} \\

If a fluid confined between two plates is brought close to its critical point, an effective force arises
acting on the walls due to the boundary conditions imposing a restriction on the spectrum of the critical 
order parameter fluctuations. Since this force has a similar origin as the Casimir force between two 
conducting plates discovered by Casimir in 1948,\cite{Casimir1948} this force is called critical Casimir 
force and was first predicted by Fisher and de Gennes \cite{Fisher1978} and is described by universal 
scaling functions.\cite{Krech1991, Krech1992, Krech1992b}
For fluids confined between two \textit{homogeneous} substrates providing symmetry breaking boundary 
conditions the universal scaling function of the critical Casimir force can be calculated analytically 
within mean field  theory.\cite{Krech1997} The Casimir forces due to director fluctuations in liquid 
crystals confined between a patterned and a homogeneous substrate have been recently studied within 
Gaussian approximation.\cite{Karimi2004}
Here we focus on the \textit{universal} features of the force generated by binary liquid mixtures close 
to their critical point and confined between two chemically inhomogeneous substrates with laterally varying 
preferences for the two species of the binary mixture. Within mean field theory we calculate numerically
the order parameter profiles and fully take into account the effect of critical adsorption.\cite{Diehl1986, 
Sprenger2005}

The article is organized as follows: In Sec.~\ref{sec:hom-sub} we recall some basic aspects of critical 
forces between homogeneous substrates in order to set the stage for the investigation of the effective 
forces emerging between inhomogeneous substrates in Sec.~\ref{sec:inhom-sub}. In Subsecs.~\ref{sec:pp}
and ~\ref{sec:pm} we study two substrates exhibiting the same and opposite patterns, respectively, 
while in Subsec.~\ref{sec:ph} we investigate the forces between one structured and one homogeneous 
substrate. After the comparison of these different cases in Subsec.~\ref{sec:compare}, we conclude in 
Subsec.~\ref{sec:ps} with the study of the normal and lateral forces appearing between two substrates 
as function of the misalignment of their surface structures. In Sec.~\ref{sec:summary} we summarize our 
results.

\section{\label{sec:hom-sub}Effective Forces between homogeneous substrates}

\subsection{Surface universality classes}
Binary liquid mixtures near their critical point of demixing belong to the so-called Ising universality 
class which also contains one-component (and multi-component) liquids near their liquid-vapor critical 
point and uniaxial ferromagnets near the Curie point. In this sense in the following we refer to binary
liquid mixtures without loss of generality.
Surfaces endow confined systems of a given bulk universality class with a fine structure of surface 
universality classes as far as their surface critical behavior is concerned. They are characterized 
within renormalization group theory by the surface enhancement $c$ and the surface field $h_1$. 
Accordingly, the bulk Ising universality class splits up into three surface universality 
classes:\cite{Diehl1986, Binder1983}
The parameter set ($h_1=0, c>0$) characterizes the so-called ordinary surface universality class for 
which the order parameter at the surface is suppressed below its bulk value. For magnetic systems this 
is the generic case.
($h_1=0, c=0$) describes (within mean field theory) the so-called special surface universality class 
with a multicritical point at which the confined system undergoes simultaneously a phase transition 
in the bulk and at the surface.
For ($h_1=0, c<0$) the system belongs to the so-called extraordinary surface universality class for 
which the order parameter is larger at the surface than in the bulk even above the critical temperature
for which the bulk order parameter vanishes. This is rather uncommon for magnetic systems mirroring 
the choice of the name. In contrast, for fluid systems an enhanced order parameter relative to the bulk
value is the generic case such that this surface universality class is then called normal surface 
universality class which is defined by ($|h_1|>0, c=0$).
The normal and the extraordinary surface universality class are equivalent and identical at their fixed 
points, ($|h_1| \rightarrow \infty, c=0$) for fluid systems and ($h_1=0, c\rightarrow -\infty$) for 
magnetic systems, and thus exhibit the same leading asymptotic critical 
behavior.\cite{Bray1977, Burkhardt1994}
In the following we shall focus on binary liquid mixtures close to their bulk critical demixing temperature 
$T_c$ and exposed to substrates with strong surface fields as representatives of the normal surface 
universality class. Our results equally hold for systems at the fixed point of the extraordinary surface 
universality class.

\subsection{Scaling of the film free energy and the solvation force}
The free energy $\Omega_{tot}$ of a binary liquid mixture confined between two parallel planar homogeneous 
substrates of macroscopic area $A$ which are separated by a distance $L$ and characterized by a surface 
field $h_1$ and $h_2$, respectively, and a surface enhancement $c_1$ and $c_2$, respectively, consists 
of a nonsingular, i.e., analytic background contribution $\Omega_{ns}$ and a part $\Omega$ with a 
singular dependence on temperature:
\begin{eqnarray}
  \Omega_{tot}(t, L, h_1, h_2, c_1, c_2) \hspace{40mm} \nonumber \\
  = \Omega_{ns}(t, L, h_1, h_2, c_1, c_2) + \Omega(t, L, h_1, h_2, c_1, c_2) \,.
\end{eqnarray}
We consider the case $t=(T-T_c)/T_c \to 0^\pm$ and $L \gg$ atomic length scales for which $\Omega$ is 
expected to exhibit universal behavior. Since we focus on the fixed point behavior 
\mbox{$(h_1 \to \pm \infty, c=0)$}, the dependences on $c_1$, $c_2$, $h_1$ and $h_2$ drop out from $\Omega$.

Both contributions to the free energy decompose into four distinct contributions:
\begin{equation} 
   \frac{\Omega (L)}{A} = L \omega_b + \omega_{s,1} + \omega_{s,2} + \omega(L) \,.
\end{equation}
The term $\omega_b$ is the bulk free energy density and $\omega_{s,1}$ and $\omega_{s,2}$ are the 
surface free energies per surface area $A$ of semi-infinite systems bounded by the surface $S_1$ and 
$S_2$, respectively. The finite-size contribution $\omega(L)$ vanishes in the limit $L \to \infty$ 
unless opposing boundary conditions create an interface in the middle of the film. Concerning this 
decomposition near $T_c$ see, e.g., Refs.~\onlinecite{Indekeu1986, Krech1991, Krech1992}; $\omega_b$ is 
independent of boundary conditions, $\omega_{s,i}$ depends on the boundary condition at $S_i$ only, 
and $\omega$ depends on \textit{both} boundary conditions. Near $T_c$ the singular contributions 
exhibit the following scaling behaviors:
\begin{equation} 
   \frac{\omega_b(t) (\xi_0^\pm)^d}{k_B\,T_c} 
   = \frac{a_b^\pm}{\alpha(1-\alpha)(2-\alpha)} \, |t|^{2-\alpha}
\end{equation}
and
\begin{eqnarray} 
   \frac{\omega_{s,i}(t) (\xi_0^\pm)^{d-1}}{k_B\,T_c} 
   = \frac{a_{s,i}^\pm}{\alpha_s(1-\alpha_s)(2-\alpha_s)} \, |t|^{2-\alpha_s} \,, \\
   i=1,2 \,, \nonumber 
\end{eqnarray}
where $\alpha$ is the critical exponent of the bulk specific heat and $\alpha_s=\alpha + \nu$ is the 
critical exponent of the surface specific heat, $\xi_0^\pm$ are the non-universal amplitudes of the 
bulk correlation length $\xi^\pm (t \to 0^\pm) = \xi_0^\pm |t|^{-\nu}$, $a_b^\pm$ are universal bulk 
amplitudes, and $a_s^\pm$ are universal surface amplitudes.

Using the hyper-scaling relation $2-\alpha = d \,\nu$\; \cite{Stanley1971, Binder1983} the singular part
of the film free energy can be rewritten as
\begin{eqnarray}
   \frac{\Omega}{k_B\,T_c\, A}
   &=& \frac{1}{L^{d-1}} \, \Bigg\{ 
       \frac{a_b^\pm}{\alpha(1-\alpha)(2-\alpha)} \, (\tilde{L}^\pm)^d \nonumber \\
   & & {} + \frac{a_{s,1}^\pm + a_{s,2}^\pm}{\alpha_s(1-\alpha_s)(2-\alpha_s)} \, (\tilde{L}^\pm)^{d-1}
       + \Theta^\pm(\tilde{L}^\pm) \nonumber
       \Bigg\} \\
   &=&
       \frac{1}{L^{d-1}} Y(\tilde{L}^\pm), \qquad \tilde{L}^\pm = \frac{L}{\xi^\pm} \,, 
\end{eqnarray}
where $\Theta^\pm(\tilde{L}^\pm)$ denote the universal scaling functions of the finite-size contribution.
Since the critical point of the film is shifted to a temperature below $T_c$ of the bulk, $Y$ is an 
analytic function at $T_c$ and thus exhibits the Taylor expansion
\begin{equation} 
   \! Y \left( \tilde{L}^\pm \to 0 \right) 
   = \sum_{i=0}^{\infty} \Delta_i^\pm(h_1=\pm\infty, h_2=\pm\infty) (\tilde{L}^\pm)^\frac{i}{\nu} \,.\!\!
   \label{eq:taylor-delta}
\end{equation}
In the case of boundary conditions $h_1 = +\infty, h_2= -\infty$ the analogue of the bulk critical
point in the film is completely eliminated (see, e.g., Ref.~\onlinecite{Schulz2005} and references therein).
Apart from the spatial dimensions and the dimensionality of the order parameter the universal coefficients
$\Delta_i^\pm$ depend on the boundary conditions applied at both substrates. 
In Eq.~(\ref{eq:taylor-delta}) the coefficients of zeroth order 
\mbox{$\Theta^\pm(0)=\Delta_0^\pm(h_1=\pm\infty, h_2=\pm\infty)=\Delta^{++}$} and
\mbox{$\Theta^\pm(0)=\Delta_0^\pm(h_1=\pm\infty, h_2=\mp\infty)=\Delta^{+-}$} are called Casimir amplitudes 
so that at $T_c$ the finite size contribution to the free energy scales as
\begin{equation} 
   \frac{\omega(L)}{k_B\,T_c} = L^{-(d-1)}\, \Delta \,, 
   \quad T=T_c \,, \quad \Delta = \Delta^{++}, \Delta^{+-} \,.
   \label{eq:finite-size-Tc}
\end{equation}
The solvation force $F_{tot}(t, L)$ between the substrates is given by 
$F_{tot}=-\frac{\partial \Omega_{tot}}{\partial L}$. Like the free energy it splits up into a nonsingular 
background $F_{ns}$ and a singular part $F(t, L)$. The singular part exhibits the scaling behavior
\begin{equation}
   \frac{F(t, L)}{k_B\,T_c\,A} \equiv f_{\perp,0}(t,L) 
   = (d-1) \, L^{-d}\, \tilde{f}^\pm_{\perp,0} (\tilde{L}^\pm) \,,
   \label{eq:scaling-force-hom}
\end{equation}
with the universal scaling function
\begin{flalign}
   & (d-1)\tilde{f}^\pm_{\perp,0} ( \tilde{L}^\pm) \nonumber \\
   & = - \Bigg\{ \frac{a_b^\pm}{\alpha(1-\alpha)(2-\alpha)} (\tilde{L}^\pm)^d
       + \tilde{L}^\pm \frac{d}{d \tilde{L}^\pm} \Theta^\pm(\tilde{L}^\pm) \nonumber \\
   &  \; \qquad {} - (d-1) \Theta^\pm(\tilde{L}^\pm)
       \Bigg\} \,.
    \label{eq:scaling-funct-force}
\end{flalign}
In the view of the analysis following later the index $0$ indicates that homogeneous substrates with
no lateral structure are considered.
One general property of the scaling function $\tilde{f}^\pm_{\perp,0}$ of the force is that for 
$T \neq T_c$ it vanishes exponentially for $L \to \infty$:\cite{Krech1991, Krech1992, Krech1992b, 
Krech1997}
\begin{equation} 
   \tilde{f}^\pm_{\perp,0} (\tilde{L}^\pm \to \infty) \sim e^{-\tilde{L}^\pm} \,.
   \label{eq:f-exponential}
\end{equation}
At $T_c$ the force is determined by the Casimir amplitude:
\begin{eqnarray} 
   f_{\perp,0}(t=0,L) = (d-1)\, L^{-d}\, \Delta_0 \,, \\[2mm]
   \Delta_0 = \Delta^{++}_0, \Delta^{+-}_0 \,. \nonumber
   \label{eq:casi-ampli}
\end{eqnarray}
The index $0$ indicates the case of homogeneous substrates.

\subsection{Scaling of the order parameter}
For laterally homogeneous surfaces $S_1$ and $S_2$ the order parameter $\phi$ depends on the normal 
distance $z$ from the surface $z=0$ and exhibits the following scaling properties:\cite{Diehl1986, 
Fisher1980, Fisher1981, Binder1983}
\begin{eqnarray}
   \phi(t, z, L) 
   &=& a |t|^\beta\, P^\pm (w^\pm, \tilde{L}^\pm), 
       \quad w^\pm=\frac{z}{\xi^\pm}, \; \tilde{L}^\pm=\frac{L}{\xi^\pm}  \nonumber \\
   &=& a \left(\frac{L}{\xi^\pm_0}\right)^{-\frac{\beta}{\nu}} \,
       \hat{P}^\pm \left( \frac{z}{L}, \tilde{L}^\pm \right) 
       \label{eq:scaling-order-1d}
\end{eqnarray}
with $\hat{P} = \left( \tilde{L}^\pm \right)^{\frac{\beta}{\nu}}P^\pm$ and $a$ as the non-universal
amplitude of the bulk order parameter. 
For increasing separation of the substrates the half-space fixed point scaling function 
$P^\pm_\infty (w^\pm)$ is attained:
\begin{equation}
   P^\pm (w^\pm, \tilde{L}^\pm \to \infty) \to P^\pm_\infty (w^\pm) \,.
\end{equation}
For a fixed scaled separation $\tilde{L}^\pm$ of the two substrates the scaling function behaves as:
\begin{equation}
   P^\pm (w^\pm \to 0, \tilde{L}^\pm) \sim (w^\pm)^{-\frac{\beta}{\nu}}
\end{equation}
and 
\begin{equation}
   P^\pm (w^\pm \to \tilde{L}^\pm, \tilde{L}^\pm) \sim (\tilde{L}^\pm-w^\pm)^{-\frac{\beta}{\nu}} \,.
\end{equation}
The half-space fixed point scaling function decays exponentially towards its bulk value $P^\pm_b$:
\begin{equation}
   P^\pm_\infty (w^\pm \to \infty) - P^\pm_b \sim e^{-w^\pm} \,,
\end{equation}
where $P^+_b=0$ and $P^-_b=1$.

\subsection{Mean field theory}
In the spirit of a systematic field theoretical renormalization group approach the scaling functions 
introduced above can be determined in lowest order perturbation theory, i.e., within mean field theory 
corresponding to $\epsilon=4-d=0$. The scaling variables entering into the scaling functions are used 
with their full scaling form for $\epsilon=1$. This approach can be extended also to the case of 
laterally inhomogeneous confining substrates which leads to analogous scaling functions with an 
enlarged set of scaling variables.
To this end we consider the fixed point Hamiltonian $\mathcal{H}[\phi]$ which is a functional 
of the order parameter profile $\phi(z)$. For binary liquid mixtures undergoing demixing it describes 
the local deviation of the concentration from the critical concentration in the bulk.
The dimensionless fixed point Hamiltonian $\mathcal{H}[\phi]= \mathcal{H}_b[\phi]+ \mathcal{H}_s[\phi]$ 
(providing the statistical weight $e^{-\mathcal{H}[\phi]}$ for the order parameter $\phi$) separates 
into the bulk part $\mathcal{H}_b[\phi]$ confined to the volume $V= A \,L$ and the surface part 
$\mathcal{H}_s[\phi]$ at the two surfaces with area $A$:\cite{Diehl1986, Binder1983}
\begin{eqnarray}
   \mathcal{H}_b [\phi] &=& \int_V \! \dd^{d-1} \mathbf{r_{\parallel}} \int_0^L\,\dd z 
                            \left( \frac{1}{2}(\nabla \phi)^2 + \frac{\tau}{2}\,\phi^2 
                                   + \frac{u}{4!}\,\phi^4 \right) \,, \nonumber \\
   \label{eq:Ham-bulk} \\
   \mathcal{H}_s [\phi] &=& \int_A \! \dd^{d-1} \mathbf{r_{\parallel}} \left( 
                                       - h_1 \phi(z=0) - h_2 \phi(z=L) \right) \,.
   \label{eq:Ham-surf}
\end{eqnarray}
Here $\mathbf{r_{\parallel}}$ is a vector parallel to the substrates, $z$ indicates the direction 
perpendicular to the substrate, and $d$ denotes the spatial dimension of the system.
$\tau$ is proportional to the reduced temperature $t=\frac{T-T_c}{T_c}$, the coupling constant $u>0$,
which is dimensionless for $d=4$, stabilizes the Hamiltonian $\mathcal{H}[\phi]$ for temperatures 
below the critical point ($T<T_c$), and $(\nabla \phi)^2$ penalizes spatial variations; $h_1$ and $h_2$ 
denote the surface fields for the substrate at $z=0$ and at $z=L$. Accordingly, the Lagrange density 
$\mathcal{L}_b[\phi]$ corresponding to the bulk part $\mathcal{H}_b[\phi]=\int \dd V \mathcal{L}_b[\phi]$ 
of the Hamiltonian reads:
\begin{equation}
   \mathcal{L}_b[\phi]= \frac{1}{2} (\nabla \phi)^2 + \frac{\tau}{2}\,\phi^2 + \frac{u}{4!}\,\phi^4 \,.
   \label{eq:Lagrange}
\end{equation}
The functional form of the Lagrange density $\mathcal{L}_s[\phi]$ of the surface part $\mathcal{H}_s[\phi]$
is independent of $L$ and leads to the boundary conditions, c.f., in Eqs.~(\ref{eq:bound-cond-sub1}) and 
(\ref{eq:bound-cond-sub2}).

Within mean field approximation fluctuations of the order parameter profile are neglected and only 
the configuration $m$ of the order parameter profile with the largest statistical weight 
$e^{-\mathcal{H}[\phi]}$ is taken into account. This mean field solution $m$ of the order parameter 
profile is determined by
\begin{eqnarray}
   \left. \frac{\delta \mathcal{H}[\phi]}{\delta \phi} \right|_{\phi = m} = 0 \,.
   \label{eq:mfa}
\end{eqnarray}

For a system confined between two parallel homogeneous substrates Eqs.~(\ref{eq:Ham-bulk})-(\ref{eq:mfa}) 
lead to a differential equation for the mean field profile $m(z)$ of the order parameter \cite{Binder1983, 
Diehl1986}
\begin{equation}
   - \partial_z^2 m + \tau \, m + \frac{u}{3!} \, m^3 = 0 \,,
   \label{eq:diff-eq} 
\end{equation}
with boundary conditions
\begin{equation}
   \left. \partial_z m\right|_{z=0} = - h_1
   \label{eq:bound-cond-sub1}
\end{equation}
and
\begin{equation}
   \left. \partial_z m \right|_{z=L} = + h_2 \,.
   \label{eq:bound-cond-sub2}
\end{equation}

\subsection{Calculation of the force from the free energy}
The effective force acting on the two parallel plates is the negative derivative of the free energy 
of the system with respect to the distance $L$ between the substrates. 
If this force between the substrates is calculated numerically from the difference in the mean field 
free energies of the systems with separation $L$ and $L+\Delta L$ between the substrates, numerical 
problems may arise due to the diverging contributions to the free energy from the surfaces for surface 
fields $h_{1,2} \to \pm \infty$ (see, c.f., Eq.~(\ref{eq:H-hom-t0})).

For systems confined between two parallel homogeneous substrates with boundary conditions 
(\ref{eq:bound-cond-sub1}) and (\ref{eq:bound-cond-sub2}) the free energy of the system per area $A$ 
and per $k_B\,T_c$ can be expressed in terms of the order parameter profile:\cite{Krech1997}
\begin{eqnarray}
   \frac{\Omega[m]}{k_B \, T_c \, A}
   &=& - \frac{2}{3}h_1\,m(0) - \frac{2}{3}h_2\,m(L)
       \label{eq:H-hom-t0} \\
   & & {}\!\!\!\! + \frac{L}{3} \left[\frac{1}{2} \left( \left. \partial_z m \right|_{z_0} \right)^2 
                            - \frac{\tau}{2}\,m(z_0)^2 - \frac{u}{4!} m(z_0)^4 
                     \right] \,, \nonumber
\end{eqnarray}
where $0 \le z_0 \le L$ denotes an arbitrarily chosen point between the two substrates. Whereas the 
first two terms in Eq.~(\ref{eq:H-hom-t0}) describe the surface free energy density per $k_B\,T_c$, 
the finite size contribution is contained in the third term in Eq.~(\ref{eq:H-hom-t0}). The alternative 
method for calculating the force between the substrates by using the stress tensor $T_{\mu\nu}$ (see, 
c.f., Subsec.~\ref{subsec:force-stresstensor}) yields that the term in square brackets in 
Eq.~(\ref{eq:H-hom-t0}) corresponds to the force between the substrates.\cite{Krech1997} In the 
following we pursue the stress tensor approach.

\subsection{\label{subsec:force-stresstensor}Force calculation using the stress tensor}
This method not only has the advantage that it avoids calculating the difference between two diverging 
free energies but also that for determining the force at a distance $L$ only the order parameter profile 
for this distance $L$ is needed.
Here we present this method for homogeneous substrates. It is applicable in the same manner for 
inhomogeneous substrates (see, c.f., Subsec.~\ref{sec:inhom-general}).

The force is a response to the change of the free energy of a system if one of the confining substrates
is shifted. A shift of one of the substrates by an infinitesimal vector 
\mbox{$\mathbf{a}(\vr)=(a_\mu), \mu=1, \ldots, d$}, leads to a displacement $\delta r_\mu$ of the components 
of the vector $\vr$, a shift $\delta m$ of the order parameter, and a change $\delta \mathcal{H}$ in 
the free energy which is given by an integral over the stress tensor $\mathcal{T}_{\mu\nu}$ and the 
improvement term $\mathcal{I}_{\mu\nu}$:\cite{Collins1976, Brown1980}
\begin{equation}
   \delta \mathcal{H} = \int \dd^d \vr \left\{ \partial_\nu a_\mu \left[ 
                             \mathcal{T}_{\mu\nu}(\vr) + \mathcal{I}_{\mu\nu}(\vr)
                        \right] \right\} \,,
   \label{eq:delta_H}
\end{equation}
where the stress tensor $\mathcal{T}_{\mu\nu}$ and the improvement term $\mathcal{I}_{\mu\nu}$ are 
given in terms of the Lagrange density $\mathcal{L}$, the order parameter $m$, and their derivatives:
\begin{eqnarray}
   \mathcal{T}_{\mu\nu} &=& \frac{\partial \mathcal{L}}{\partial (\partial_\nu m)} \partial_\mu m 
                            - \delta_{\mu\nu} \mathcal L \,,
   \label{eq:T_munu} \\
   \mathcal{I}_{\mu\nu} &=& - \frac{1}{4} \frac{d-2}{d-1}
                             (\partial_\nu \partial_\mu - \delta_{\mu\nu} \Delta) m^2 \,,
   \label{eq:I_munu}
\end{eqnarray}
where $\delta_{\mu\nu}$ is the Kronecker symbol.

Now we consider a system confined between two homogeneous substrates and described by the Lagrange 
density in Eq.~(\ref{eq:Lagrange}). For this system within mean field theory, i.e., $d=4$, only the 
component $\mathcal{T}_{zz}$ of the stress tensor is nonzero:
\begin{equation}
   \mathcal{T}_{zz} = \frac{1}{2}(\partial_z m)^2 - \frac{\tau}{2}\,m^2 - \frac{u}{4!}\,m^4 \,.
   \label{eq:stress-tensor-hom}
\end{equation}
If one of the substrates is shifted by $\alpha$ in normal direction ($z$-direction), 
$\mathbf{a}(\vr) = \alpha \theta(z-z_0) \mathbf{e}_z$, where $z_0$ is an arbitrary position between 
the substrates and $\theta$ is the Heaviside step function, the Hamiltonian changes by (in the laterally
homogeneous case $\mathcal{I}_{\mu\nu}=0$)
\begin{eqnarray}
   \delta \mathcal{H} &=& \int \dd ^d \vr \left[ (\partial_z a_z) T_{zz} \right]
   \label{eq:Ham-hom-sub-Tzz} \\
                      &=& \int \dd^{d-1} \vr_{\parallel} \
                               \int \dd z \left( \alpha \delta(z-z_0) \mathcal{T}_{zz} \right) \nonumber \\
                      &=& A\, \alpha \, \mathcal{T}_{zz} (z=z_0) \,. \nonumber
\end{eqnarray}
Within mean field theory ($d=4$), $A$ is a three-dimensional surface area. 
This change $\delta \mathcal{H}$ results in a force $F_{\perp,0}$ in $z$-direction between the two 
substrates:
\begin{eqnarray}
   \frac{F_{\perp,0}}{k_B\,T_c} &=& \frac{\delta \mathcal{H}}{\partial \alpha}
   \label{eq:force_zz-h_zz} \\
                                &=& A\, \mathcal{T}_{zz} (z=z_0) \,, \nonumber
\end{eqnarray}
where in view of the analysis following below the index $0$ indicates that homogeneous substrates with 
no lateral structure are considered. For $F_{\perp,0}<0$ ($>0$) the confining substrates attract (repel) 
each other.
With Eqs.~(\ref{eq:stress-tensor-hom}) and (\ref{eq:force_zz-h_zz}) the force 
$f_{\perp,0} = \frac{F_{\perp,0}}{k_B\,T_c\,A}$ per area and $k_B\,T_c$ is given by
\begin{equation}
   f_{\perp,0}(t,L) = \frac{1}{2}(\partial_z m(z_0))^2 - \frac{\tau}{2}\,m(z_0)^2 
                       - \frac{u}{4!}\,m(z_0)^4 \,.
   \label{eq:force_final}
\end{equation}
Equation~(\ref{eq:force_final}) is equivalent to the term in the square brackets in 
Eq.~(\ref{eq:H-hom-t0}). As expected the force $f_{\perp,0}(L)$ exhibits the scaling property described 
in Eq.~(\ref{eq:scaling-force-hom}) with $\xi^\pm = \xi^\pm_0 |t|^{-\nu}$ where $\xi^+ = 1/\sqrt{\tau}$ 
and $\xi^- = 1/\sqrt{2|\tau|}$. 
According to Eq.~(\ref{eq:force_final}) the mean field expressions for $f_{\perp,0}$ and 
$\tilde{f}_{\perp,0}$ contain a common prefactor $6/u$ which diverges $\sim 1/\epsilon$
for $\epsilon \to 0$ if one inserts for $u$ its fixed point value $u^* \sim \epsilon$ (compare 
Eq.~(\ref{eq:scaling-funct-force}) where $\alpha \sim \epsilon$). For the quantities studied
below we shall consider ratios of such scaling functions for which this prefactor drops out so that
these ratios have finite and well defined mean field values in the limit $\epsilon \to 0$. Therefore
the mean field results for these ratios are expected to provide reasonable estimates for their actual
counterparts in $d=3$. \\

In the case of homogeneous substrates the force between the substrates can be calculated directly
based on the stress tensor without calculating the order parameter profile $m(z)$.\cite{Krech1997}
For the fixed point values $h_1 \to +\infty$, $h_2 \to +\infty$, $c_1=c_2=0$ the scaling function 
$\tilde{f}^{++}_{\perp,0}$ of the force (the index pair $++$ denotes that the two surface fields are 
parallel; the symbol $\pm$ for $T \lessgtr T_c$ is suppressed) is given in terms of the complete 
elliptic integral of the first kind 
$K(k)=\int_0^{\pi/2} \frac{\dd\varphi}{\sqrt{1-k^2\sin^2\varphi}}$, with the modulus $k \in [0,1)$:
\begin{flalign}
   & \tilde{f}^{++}_{\perp,0} (\tilde{L}^+) \nonumber \\
   & = \frac{6}{(d-1)u} \Bigg\{ (2K(k))^4 k^2 (k^2-1) 
      - \theta(-t) \, \frac{(\tilde{L}^+)^4}{4} \Bigg\} \,, \nonumber \\
   & t > t^* = - \left( \frac{\pi \xi^+_0}{L} \right)^2 ,
     \label{eq:t_star}
\end{flalign}
where $k=k(\tilde{L}^+)$ is determined implicitly via
\begin{equation}
   \tilde{L}^+ (t > t^*) = \sqrt{ |(2K(k))^2 (2k^2-1)| } \,. 
\end{equation}
The term $\frac{(\tilde{L}^+)^4}{4}$ is the bulk contribution to the force below the bulk critical 
temperature. (If one considers the slit system together with the thin walls to be immersed 
into the bulk liquid, this bulk contribution to the effective force is absent.) For simplicity and
practical advantages below we keep the notation $\tilde{L}^+$ also for temperatures \mbox{$t<0$}, 
with the meaning $\tilde{L}^+(t < t^*) = \tilde{L}^- \xi_0^-/\xi_0^+$.

\begin{figure}
   \begin{center}
      \epsfig{file=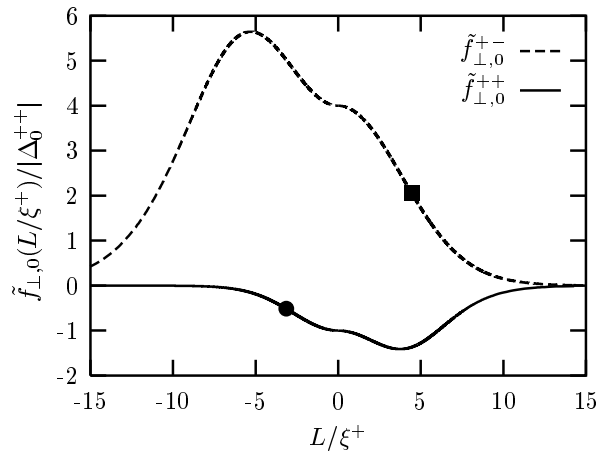,width=0.48\textwidth}
      \caption{\label{fig:force-hom_sub}Universal scaling functions $\tilde{f}^{++}_{\perp,0}(\tilde{L})$  
               and $\tilde{f}^{+-}_{\perp,0}(\tilde{L})$ of the force within mean field theory (see 
               Eqs.~(\ref{eq:scaling-force-hom}) and (\ref{eq:force_final})) between two parallel  
               geometrically flat and chemically homogeneous substrates at distance $L$ for strong 
               parallel $(++)$ and antiparallel $(+-)$ surface fields $h_1$ and $h_2$. For comparison
               both scaling functions are normalized by the absolute value of $\tilde{f}^{++}_{\perp,0}(0)$
               at $T_c$ given by $|\Delta_0^{++}|$ (see Eq.~(\ref{eq:casimir-pp})). Note that in order
               to avoid breaks in the slope for the curves here we have used the same scaling variable
               $\tilde{L}^+=L/\xi^+$ both above ($\tilde{L}>0$) and below ($\tilde{L}<0$) $T_c$. 
               {\Large $\bullet$} and $\blacksquare$ correspond to $L/\xi(t=t^*)$ (Eq.~(\ref{eq:t_star})) and
               $L/\xi(t=t^\circ)$ (Eq.~(\ref{eq:t_circ})), respectively.} 
   \end{center}   
\end{figure}

For the temperature range $t < t^*$ one has
\begin{flalign}
   & \tilde{f}^{++}_{\perp,0} (\tilde{L}^+)
     = \frac{6}{(d-1)u} \left\{
     (2K(k))^4 k^2 - \theta(-t)  \, \frac{(\tilde{L}^+)^4}{4} \right\} \,, \nonumber \\ 
   & t < t^* \,,
\end{flalign}
with
\begin{equation}
   \tilde{L}^+ (t < t^*) = \sqrt{ (2K(k))^2 (k^2+1) } \,.
\end{equation}

In the case of strong antiparallel surface fields 
$h_1 \to + \infty$, $h_2 \to -\infty$ one has \cite{Krech1997}
\begin{flalign}
   & \tilde{f}^{+-}_{\perp,0} (\tilde{L}^+) \nonumber \\
   & = \frac{6}{(d-1)u} \left\{
     (2K(k))^4(1-k^2)^2 - \theta(-t) \, \frac{(\tilde{L}^+)^4}{4} \right\}\,, \nonumber \\
   & t>t^\circ = 2 \left( \frac{\pi \xi^+_0}{L} \right)^2 ,
   \label{eq:t_circ}
\end{flalign}
with 
\begin{equation}
   \tilde{L}^+ (t > t^\circ) = \sqrt{ 2 (2K(k))^2 (k^2+1) } \,.
\end{equation}

For the temperature range $t < t^\circ$ one has 
\begin{flalign}
   & \tilde{f}^{+-}_{\perp,0} (\tilde{L}^+)
      = \frac{6}{(d-1)u} \left\{
     (2K(k))^4 - \theta(-t) \, \frac{(\tilde{L}^+)^4}{4} \right\} \,, \nonumber \\
   & t < t^\circ \,,
\end{flalign}
with
\begin{equation}
   \tilde{L}^+ (t < t^\circ) = \sqrt{ |2 (2K(k))^2 (2k^2-1)| } \,.
\end{equation}
Figure~\ref{fig:force-hom_sub} shows these scaling functions.

For parallel surface fields the force $\tilde{f}^{++}_{\perp,0}(\sign(t)\tilde{L}^+)$ is always negative, 
i.e., attractive, and exhibits a minimum at a temperature above $T_c$. 
For antiparallel surface fields the force $\tilde{f}^{+-}_{\perp,0}(\sign(t)\tilde{L}^+)$ is positive, 
i.e., repulsive. For $\tilde{L}^+ \to \infty$ the forces vanish $\sim e^{-\tilde{L}^+} = e^{-L/\xi^+}$ 
for $t>0$ whereas they vanish $\sim e^{-\tilde{L}^+ \xi^+_0/\xi^-_0} = e^{-\tilde{L}^-} = e^{-L/\xi^-}$ 
for $t<0$.

At the critical point, the forces are proportional to the Casimir amplitudes, which are given by the 
scaling functions at $\tilde{L}=0$ (see Eq.~(\ref{eq:casi-ampli})):
\begin{eqnarray}
   \tilde{f}^{++}_{\perp,0}(\tilde{L}=0) &=& \Delta^{++}_0 \,, 
   \label{eq:casimir-pp} \\
   \tilde{f}^{+-}_{\perp,0}(\tilde{L}=0) &=& \Delta^{+-}_0 \,.
   \label{eq:casimir-pm}
\end{eqnarray}
The forces at the critical point are given for $k=1/\sqrt{2}$ and for $d=4$ amount to 
\begin{eqnarray}
   \Delta^{++}_0 
   = -4 \left( K \textstyle{ \left( \frac{1}{\sqrt{2}} \right) } \right)^4 \, \frac{6}{(d-1)u} 
   = -15.7561\, \frac{6}{u}
   \label{eq:delta_pp-numbers}
\end{eqnarray} 
and 
\begin{eqnarray}
    \Delta^{+-}_0 
    = 16 \left( K \textstyle{ \left( \frac{1}{\sqrt{2}} \right) } \right)^4 \, \frac{6}{(d-1)u}
    = 63.0242\, \frac{6}{u} \,,
\end{eqnarray}
i.e., within mean field theory at the critical point the absolute value of the force between 
homogeneous substrates with antiparallel infinite surface fields is four times larger than for the 
case with parallel infinite surface fields: $\Delta^{+-}_0 = 4 |\Delta^{++}_0|$.

\section{\label{sec:inhom-sub}Forces between inhomogeneous substrates}

\subsection{\label{sec:inhom-general}Geometry and general expressions}
\begin{figure}
   \begin{center}
      \epsfig{file=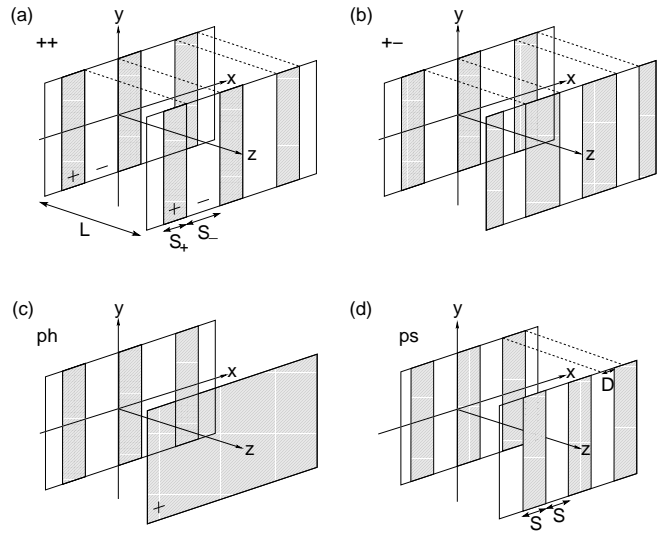,width=0.48\textwidth}
      \caption{\label{fig:systems}Two parallel, geometrically flat, and chemically inhomogeneous 
               substrates at distance $L$. We consider four generic configurations: (a) $++$, (b) $+-$, 
               (c) periodically structured substrate vs. homogeneous substrate, and (d) periodic 
               structures shifted relative to each other. On the white stripes $h_1=-\infty$, on the
               grey stripes $h_1=+\infty$.}
   \end{center}
\end{figure}
In this section we consider systems above and at the bulk critical temperature. To simplify the notations
in the following we omit the index $+$ indicating quantities above the critical point, e.g., 
$\tilde{L}^+$ is replaced by \mbox{$\tilde{L} = L/\xi^+ = L / \xi$}.

The substrates investigated here exhibit inhomogeneities in one lateral direction ($x$~direction) and
translational invariance in the remaining $d-2$ directions $y_1$, \ldots, $y_{d-2}$ (extension $H$). 
The lateral substrate inhomogeneities form periodic patterns of "positive" stripes ($h_1=+\infty$) of 
scaled width $\tilde{S}_+= S_+ / \xi$ and "negative" stripes (\mbox{$h_1=-\infty$}) of scaled width 
$\tilde{S}_- = S_- / \xi$ with periodic boundary conditions in $x$~direction (see 
Fig.~\ref{fig:systems}). (In the preceding paper \cite{Sprenger2005} these quantities are denoted 
as $\tilde{S}_p$ and $\tilde{S}_n$, respectively.) In the following four different combinations 
of the substrate structures are investigated:
\begin{enumerate}
  \item $++$~configuration: 
    two identically patterned substrates with parallel alignment of the infinite surface fields on the
    opposing, unshifted stripes.
  \item $+-$~configuration:
    two identically patterned substrates with antiparallel alignment of the infinite surface fields
    on the opposing, unshifted stripes.
  \item $ph$~configuration (\textit{"periodic-homogeneous"} configuration):
    a patterned substrate opposing a homogeneous substrate.
  \item $ps$~configuration (\textit{"periodic-shifted"} configuration):
    two substrates with the same periodic pattern of stripes of width $S$ but with varying shift $D$ 
    of the stripes.
\end{enumerate}
Various length scales enter describing the inhomogeneities of the substrates, i.e., the set 
$(L ,S_+, S_-)$ in the cases of the $++$~configuration, the $+-$~configuration, and the 
$ph$~\textit{con}figuration. The scaling function $\tilde{f}^{con}(\tilde{L})$ of the force (with 
$con=(++, +-, ph)$) between the two patterned substrates depends on the scaled distance $\tilde{L}$ 
and the scaled stripe widths $\tilde{S}_+ = S_+/\xi$ and $\tilde{S}_- = S_-/\xi$ or equivalently 
on the set ($\tilde{L}$, $S_+/L$, $S_-/S_+$). The latter set exhibits the advantage that upon approaching 
the critical temperature ($t \to 0$, $\xi \to \infty$) only the variable $\tilde{L}$ vanishes. The 
scaling behavior of the normal force between two inhomogeneous substrates is given by (compare 
Eq.~(\ref{eq:scaling-force-hom}))
\begin{eqnarray}
   \frac{F^{con}_\perp(t,L, S_+, S_-)}{k_B\,T_c\,A} 
   &=& f^{con}_\perp(t, L, S_+, S_-) 
   \label{eq:scaling-force-inhom-con} \\
   &=& (d-1)\, L^{-d} \, \tilde{f}^{con}_\perp(\tilde{L}, S_+/L, S_-/S_+) \,. \nonumber 
\end{eqnarray}
Without lateral shifts for these configurations $(++, +-, ph)$ the lateral force $F^{con}_\parallel$
is zero.

For the $ps$~configuration the parameters entering are the separation $L$ between the substrates, the 
stripe width $S$, and the distance $D$ by which the stripes of the second substrate are shifted relatively
to the first substrate. The scaling function $\tilde{f}^{ps}(\tilde{L})$ of the force is determined
by the set $(\tilde{L}, \tilde{S}/ \tilde{L}, \delta)$, with $\delta = \tilde{D}/\tilde{S}$. In this case
the scaling behavior of the force normal and lateral to the substrates, respectively, is given by (compare 
Eq.~(\ref{eq:scaling-force-hom})):
\begin{eqnarray}
   \frac{F^{ps}_\perp(t, L, S, D)}{k_B\,T_c\,A} 
   &=& f^{ps}_\perp(t, L, S, D) \\
   &=& (d-1)\, L^{-d} \, \tilde{f}^{ps}_\perp(\tilde{L}, S/L, \delta) \nonumber
\end{eqnarray}
and 
\begin{eqnarray}
   \frac{F^{ps}_\parallel(t, L, S, D)}{k_B\,T_c\,A} 
   &=& f^{ps}_\parallel(t ,L, S, D) \\
   &=& (d-1)\, L^{-d} \, \tilde{f}^{ps}_\parallel(\tilde{L}, S/L, \delta) \,. \nonumber
   \label{eq:scaling-force-inhom-ps}
 \end{eqnarray}
For the \textit{con}figurations $++$, $+-$ and $ph$ the order parameter profile exhibits the scaling behavior
(with \mbox{$con=++, +-, ph$}):
\begin{flalign}
   & \phi(t, x, z, L, S_+, S_-) \nonumber \\
   & = a |t|^{\beta} P^{\mathit{con}} \left( 
     v=\frac{x}{\xi}, w=\frac{z}{\xi}, \tilde{L}=\frac{L}{\xi}, \frac{S_+}{L}, \frac{S_-}{S_+} \right)
\end{flalign}
where $P^{\mathit{con}}(v,w,\tilde{L}, S_+/L, S_-/S_+)$ is a universal scaling function and $a$ the 
non-universal amplitude of the bulk order parameter (compare Eq.~(\ref{eq:scaling-order-1d})).
In the $ps$ case, the scaling variables $S_+/L$ and $S_-/S_+$ are again replaced by $S/L$ and $\delta$.

Quantitative results for the scaling functions of the forces follow from the fixed-point Hamiltonian 
\mbox{$\mathcal{H}[\phi]= \mathcal{H}_b[\phi]+ \mathcal{H}_s[\phi]$} (compare Eqs.~(\ref{eq:Ham-bulk}) 
and (\ref{eq:Ham-surf})). Here the integration over the surface area $A=N\,H\,B$ is
$\int \dd^{d-1} \vr_\parallel = N \int_0^H \dd^{d-2}\vr_\parallel \int_0^B \dd x$, with $B=2S$ for 
the $ps$~configuration and $B=S_+ + S_-$ for the other configurations; $N$ is the number of repeat 
units in $x$ direction.

The mean field approximation $m(x,z)$ of the order parameter profile fulfills the partial differential 
equation
\begin{eqnarray}
   - \left( \partial_x^2 + \partial_z^2 \right)m + \tau m + \frac{u}{3!} m^3 = 0 \,.
   \label{eq:diff-eq-2d} 
\end{eqnarray}
The accompanying boundary conditions are given by Eqs.~(\ref{eq:bound-cond-sub1}) and 
(\ref{eq:bound-cond-sub2}) with $h_1$ and $h_2$ varying along $x$ steplike at the chemical steps on
the substrate.

From the order parameter profile the force between the substrates can be calculated via the stress 
tensor method (see Subsec.~\ref{subsec:force-stresstensor}). The inhomogeneities in $x$~direction 
lead to further non-vanishing components of the stress tensor $\mathcal{T}_{\mu\nu}$ and of the 
improvement term $\mathcal{I}_{\mu\nu}$ (see Eqs.~(\ref{eq:Lagrange}), (\ref{eq:T_munu}) and 
(\ref{eq:I_munu})).

A shift of one substrate by $\alpha$ in the $z$-direction, 
\mbox{$\mathbf{a}(\vr) = \alpha \theta(z-z_0) \mathbf{e}_z$}, where $z_0$ is an arbitrary position 
between the substrates, yields the normal force $f_\perp(L)$ per $k_B\,T_c$ and per area 
\mbox{$A=N\,B\,H$}:
\begin{equation}
   \frac{F_\perp}{k_B\,T_c\,N\,B\,H}
   = f_\perp(L) = \frac{1}{B} \int_0^B \dd x \, (\mathcal{T}_{zz}(z_0) + \mathcal{J}_{zz}(z_0)) \,,
   \label{eq:stress-methode-perp}
\end{equation}
with the components (recall that $\partial_{y_i}m=0$)
\begin{equation}
   \mathcal{T}_{zz} = - \frac{1}{2} (\partial_x m)^2 + \frac{1}{2}(\partial_z m)^2 
                        - \frac{\tau}{2} \, m^2 - \frac{u}{4!} \, m^4
\end{equation}
and 
\begin{equation}
   \mathcal{J}_{zz} = \frac{1}{3}(\partial_x m)^2 + \frac{1}{3} m \, \partial_x^2 m \,.
   \label{eq:stress-improv-kompo-perp}
\end{equation}

The lateral force $f_\parallel(L)$ per $k_B\,T_C$ and per area \mbox{$A=N\,B\,H$} follows from considering 
a shift by $\beta$ in \mbox{$x$-direction}, $\mathbf{b}(\vr) = \beta \theta(z-z_0) \mathbf{e}_x$:
\begin{equation}
   \frac{F_\parallel}{k_B\,T_c\,N\,B\,H} = {f}_\parallel(L) 
    = \frac{1}{B} \int_0^B \dd x \, (\mathcal{T}_{xz}(z_0) + \mathcal{J}_{xz}(z_0)) \,,
   \label{eq:f_xz}
\end{equation}
with
\begin{equation}
   \mathcal{T}_{xz} = (\partial_x m) \, (\partial_z m)
\end{equation}
and 
\begin{equation}
   \mathcal{J}_{xz} = - \frac{1}{3} (\partial_x m) \, (\partial_z m)
                        - \frac{1}{3} m \, \partial_x \partial_z m \,.
   \label{eq:stress-improv-kompo-para}
\end{equation}
Within the present mean field level ($d=4$) the scaling functions $\tilde{f}_\perp(\tilde{L})$ and 
$\tilde{f}_\parallel(\tilde{L})$ as defined in Eqs.~(\ref{eq:scaling-force-inhom-con}) - 
(\ref{eq:scaling-force-inhom-ps}) can be expressed in terms of the corresponding scaling functions 
$P$ of the order parameter:
\begin{eqnarray}
   \tilde{f}_\perp(\tilde{L}) 
   &=& \frac{6}{(d-1)u} \frac{\tilde{L}^4}{\tilde{B}} \times \\
   & & \int_0^{\tilde{B}} \dd v \left( 
           - \frac{1}{6}(\partial_v \left.P\right|_{w_0})^2 
           + \frac{1}{2}(\partial_w \left.P\right|_{w_0})^2 \right. \nonumber \\
   & & \left. {} + \frac{1}{3} P(w_0)\partial_v^2 \left.P\right|_{w_0}
                 - \frac{1}{2} P(w_0)^2 - \frac{1}{4} P(w_0)^4 
       \right) \,, \nonumber \\
   \tilde{f}_\parallel(\tilde{L})
   &=& \frac{6}{(d-1)u} \frac{\tilde{L}^4}{\tilde{B}} \int_0^{\tilde{B}} \dd v \left( 
           \frac{2}{3} \partial_v \left.P\right|_{w_0} \partial_w  \left.P\right|_{w_0} 
        \right. \nonumber \\
   & & \left. \hspace{23mm} {} - \frac{1}{3} P(w_0) \partial_v \partial_w \left.P\right|_{w_0}  
      \right) \,.
   \label{eq:f_vw-P}
\end{eqnarray}
Here $\tilde{B}=B/ \xi^+$ and $\tilde{L}= L/ \xi^+$ denote the scaled extensions of the system in 
the $x$ and the $z$ direction, respectively.

At the critical point, i.e., for $\tilde{L}=0$ the scaling function $\tilde{f}_\perp$ of the force 
reduces to a generalized Casimir amplitude $\Delta$ which depends on the type of the inhomogeneities 
of the substrates:
\begin{eqnarray}
   \tilde{f}_\perp(\tilde{L}=0) &=& \Delta^{\mathit{con}} \qquad con=++, +-, ph, ps \,.
\end{eqnarray}

\subsection{\label{sec:pp}Forces for the $++$~configuration}
In this subsection we consider two identical parallel flat substrates with the relative positions of
the patterns being in phase (see Fig.~\ref{fig:systems}(a)).

First we want to state some general properties of the force $\tilde{f}^{++}_\perp$. The force 
$\tilde{f}^{++}_\perp$ is symmetric w.r.t. to interchanging the width $\tilde{S}_+$ of the positive 
and the width $\tilde{S}_-$ of the negative stripes:
\begin{eqnarray}
  \tilde{f}^{++}_\perp (\tilde{L},q_1, q_2) 
  = \tilde{f}^{++}_\perp \left( \tilde{L},q_1q_2, \frac{1}{q_2} \right)\,, \\
  q_1=\frac{S_+}{L}, \; q_2=\frac{S_-}{S_+} \,. \nonumber
  \label{eq:fpp-symmetry-Sp-Sn}
\end{eqnarray}
This symmetry equally holds in the case of the $+-$~configuration for the scaling function of the force 
$\tilde{f}^{+-}_\perp$ (see, c.f., Subsec.~\ref{sec:pm}).

\begin{figure}
   \begin{center}
      \epsfig{file=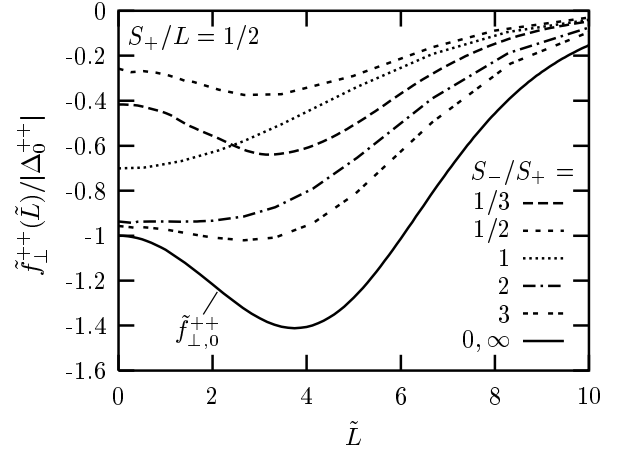,width=0.48\textwidth}
      \caption{\label{fig:pp-SpL1_2-SnSpx}Scaling function $\tilde{f}^{++}_\perp(\tilde{L})$ of the 
               force between two equally patterned substrates, which are in phase, as a function of
               the scaled  distance $\tilde{L}=L/\xi$ between the substrates for different ratios 
               $S_-/S_+$ of the stripe widths (dashed curves) and for a fixed ratio $S_+/L=1/2$ 
               compared with the scaling function $\tilde{f}^{++}_{\perp,0}(\tilde{L})$ of the force 
               between two homogeneous substrates (solid curve). The curves are normalized by the absolute 
               value of the Casimir amplitude $|\Delta^{++}_0|$ for the case of homogeneous substrates 
               (see Eq.~(\ref{eq:casimir-pp})). The surface fields on opposing stripes are pointing into 
               the same direction.} 
   \end{center}
\end{figure}

If one fixes the width of the positive stripe $\tilde{S}_+$ and considers the limits $\tilde{S}_- \to 0$
or $\tilde{S}_- \to \infty$, the force reduces to that between two corresponding homogeneous substrates:
\begin{eqnarray}
  \tilde{f}^{++}_\perp \left( \tilde{L}, \frac{\tilde{S}_+}{\tilde{L}}, 
                              \frac{\tilde{S}_-}{\tilde{S}_+} \to 0 \right) 
  &=& \tilde{f}^{++}_{\perp,0}(\tilde{L}) 
  \label{eq:fpp-Sn-null} \\
  \tilde{f}^{++}_\perp \left( \tilde{L}, \frac{\tilde{S}_+}{\tilde{L}}, 
                              \frac{\tilde{S}_-}{\tilde{S}_+} \to \infty \right)
  &=& \tilde{f}^{--}_{\perp,0}(\tilde{L}) \nonumber \\
  &=& \tilde{f}^{++}_{\perp,0}(\tilde{L})
  \label{eq:fpp-Sn-inf}
\end{eqnarray}
This can be seen in Fig.~\ref{fig:pp-SpL1_2-SnSpx}.

\begin{figure}
   \begin{minipage}{0.48\textwidth}
   \begin{center}
      \epsfig{file=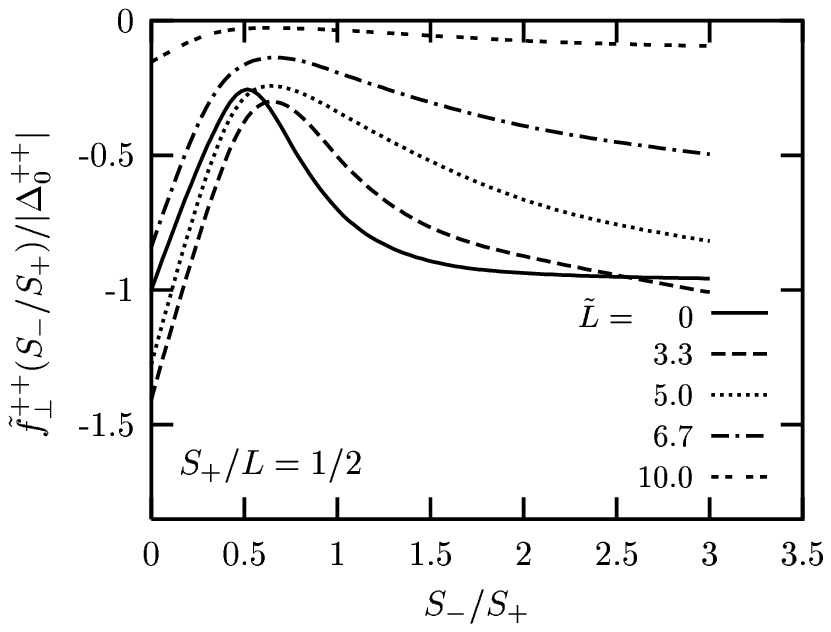,width=\textwidth}
      \caption{\label{fig:pp-SnSp-L}Scaling function $\tilde{f}^{++}_\perp(S_-/S_+)$ of the force 
               between two substrates with their surface field patterns in phase and normalized by 
               $|\Delta^{++}_0|$ as a function of the ratio $S_-/S_+$ of the stripe widths for various 
               fixed distances $\tilde{L}$ between the substrates and a ratio $S_+/L = 1/2$. The influence
               of the ratio $S_-/S_+$ on the force becomes weaker with increasing distance. For 
               $S_-/S_+ = 0, \infty$ one recovers the scaling function for the force between two  
               homogeneous substrates, i.e., for each curve the values at $S_-/S_+ = 0$ and 
               $S_-/S_+ = \infty$ are equal. The solid line represents the system at $T_c$, i.e., for 
               $\tilde{L}=0$; in this case for large $L$ with $S_+/L = const$ and $S_-/S_+ \to 0$ or 
               $S_-/S_+ \to \infty$ one recovers the Casimir amplitude $\Delta^{++}_0$ for the homogeneous
               case so that due to the normalization the solid curve reaches $-1$.}
      \epsfig{file=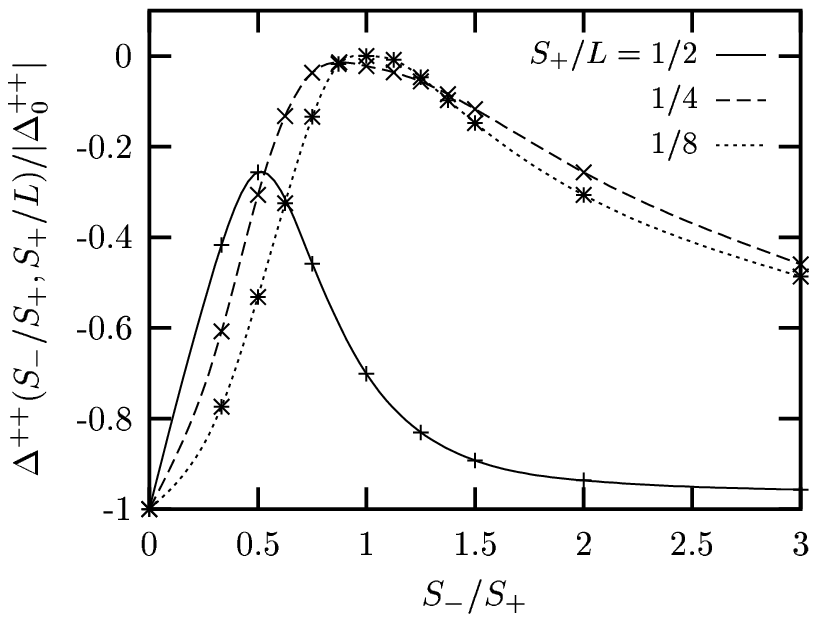,width=\textwidth}
      \caption{\label{fig:casimir-pp-SnSp}Generalized Casimir amplitude $\Delta^{++}(S_-/S_+, S_+/L)$ 
               for a critical fluid confined between two substrates with their surface field patterns in 
               phase and normalized by $|\Delta^{++}_0|$ as a function of the stripe width ratio $S_-/S_+$ 
               for different ratios $S_+/L$. The curve for $S_+/L=1/8$ serves as an approximation for 
               the actual generalized Casimir amplitude $\Delta^{++}(S_-/S_+)$ corresponding to the 
               limit $S_+/L \to 0$. $\Delta^{++}(S_-/S_+)$ varies between $\Delta^{++}_0<0$ for 
               $S_-/S_+ = 0, \infty$ and its maximum value $(\Delta^{++})_{\mathit{max}} \simeq 0$ 
               (see main text). Here and in the following figures the curves are smoothly interpolating 
               between the indicated data points.}
   \end{center}
   \end{minipage}
\end{figure}

At the critical temperature $T_c$, i.e., for $\tilde{L}=0$, the scaling function of the force depends 
on $S_+/L$ and $S_-/S_+$. In the subsequent limit of large substrate separations (i.e., for 
\mbox{$S_+/L \to 0$}) the dependence on $S_+/L$ drops out so that asymptotically the force is governed 
by a generalized Casimir amplitude $\Delta^{++}(S_-/S_+)$ (see, c.f., Fig.~\ref{fig:casimir-pp-SnSp}).

Figure~\ref{fig:pp-SpL1_2-SnSpx} shows the dependence of the normal force 
$\tilde{f}^{++}_\perp(\tilde{L}, S_+/L, S_-/S_+)$ on the scaled distance $\tilde{L}$ between the 
substrates for different ratios $S_-/S_+$ and a fixed ratio $S_+/L$ in comparison with the force 
$\tilde{f}^{++}_{\perp,0}(\tilde{L})$ (see Fig.~\ref{fig:force-hom_sub}) for two homogeneous substrates 
with parallel surface fields.
The strength of the attractive force $\tilde{f}^{++}_\perp(\tilde{L})$ between patterned substrates is 
reduced in comparison to the strength of the force $\tilde{f}^{++}_{\perp,0}(\tilde{L})$ between homogeneous
substrates.
At the critical point, i.e., for $\tilde{L}=0$, the force $\tilde{f}^{++}_\perp(\tilde{L})$ between 
the substrates takes a finite negative value, i.e., the force between the substrates is attractive.
The opposite limit $\tilde{L} \to \infty$ can be realized either by increasing the distance $L$ between 
the substrates or by raising the temperature. In both cases the force $\tilde{f}^{++}_\perp(\tilde{L})$ 
between the substrates is expected to vanish exponentially (compare Eq.~(\ref{eq:f-exponential})) which 
is confirmed by our data.
For a given distance $\tilde{L}$ the scaled width $\tilde{S}_+$ of the positive stripe is fixed (due to the 
choice $S_+/L = 1/2$ in Fig.~\ref{fig:pp-SpL1_2-SnSpx}) and therefore for $S_-/S_+ \to \infty$ the 
negative stripes become larger and the force between the substrates approaches the force 
$\tilde{f}^{--}_{\perp,0}(\tilde{L}) = \tilde{f}^{++}_{\perp,0}(\tilde{L})$ between two homogeneous 
substrates (see Eq.~(\ref{eq:fpp-Sn-inf})). 
With increasing stripe width the order parameter distributions at the chemical steps forming the 
stripes become more independent from each other so that the healing of the order parameter profile 
becomes more pronounced, i.e., the flattening of the cross section at $z=const$ of the order parameter 
profile across a chemical step becomes flatter with increasing distance from the 
substrate.\cite{Sprenger2005} (For narrow stripes this healing is reduced due to the mutual influence 
of the chemical steps onto each other.) Therefore one can conclude that healing of the order parameter 
distribution reduces the absolute value of the attractive force. 
In the limit $S_-/S_+ \to 0$ the negative stripes disappear and the system again approaches the 
homogeneous case (see Eq.~(\ref{eq:fpp-Sn-null})). Therefore in Fig.~\ref{fig:pp-SpL1_2-SnSpx} the
curve for $S_-/S_+ = 1/3$ lies below that one for  $S_-/S_+ = 1/2$. For the latter ratio the force is
weakened most.

\begin{figure}
   \begin{center}
      \epsfig{file=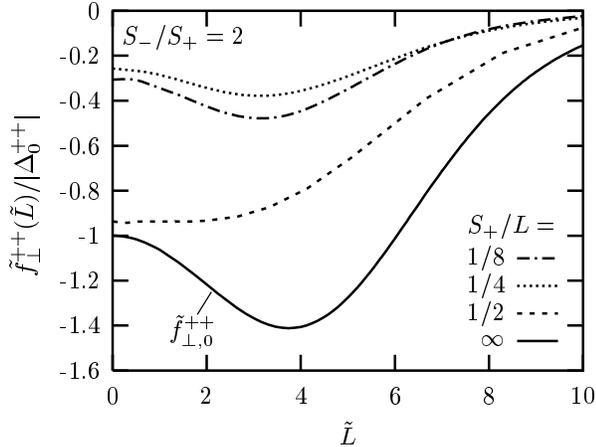,width=0.48\textwidth}
      \caption{\label{fig:pp-SpLx-SnSp2_1}Scaling function $\tilde{f}^{++}_\perp(\tilde{L})$ of the 
               force between two patterned substrates with parallel surface fields normalized by 
               $|\Delta^{++}_0|$ for different ratios $S_+/L$ and a fixed ratio $S_-/S_+=2$.}
   \end{center}
\end{figure}

The force $\tilde{f}^{++}_\perp(\tilde{L})$ between the substrates also depends on the scaled stripe 
widths $\tilde{S}_+$ and $\tilde{S}_-$ relative to the plate separation $\tilde{L}$. This is illustrated 
in Fig.~\ref{fig:pp-SpLx-SnSp2_1} for systems with a fixed ratio $S_-/S_+$ of the stripe widths but 
with different ratios $S_+/L$. It turns out that for fixed distances $\tilde{L} \gtrsim 7$ the magnitude
of the force increases for increasing ratios $S_+/L$ because this implies larger stripes. For smaller 
distances $\tilde{L}$ this trend is again partly inverted due to healing effects such that the magnitude 
of the force is no longer smallest for the smallest ratio $S_+/L$. \\

The influence of the stripe width ratio $S_-/S_+$ becomes weaker with increasing scaled distance 
$\tilde{L}$ between the substrates (see Fig.~\ref{fig:pp-SnSp-L}). In the limits $S_-/S_+ \to 0$ and 
$S_-/S_+ \to \infty$ the curves reach the values for the force $\tilde{f}^{++}_{\perp,0}(\tilde{L})$ 
between two homogeneous substrates. 
For $\tilde{L}=0$ Fig.~\ref{fig:pp-SnSp-L} renders a universal scaling function 
$\Delta^{++}(S_-/S_+, S_+/L)$ which in the limit of large $L$, \mbox{$S_-/S_+ \to 0$}, and $S_+/L=const$
or in the limit of large $L$, $S_-/S_+ \to \infty$, and $S_+/L=const$ reduces to the Casimir amplitude
$\Delta^{++}_0$ for homogeneous systems. In order to obtain the generalized Casimir amplitude 
$\Delta^{++}(S_-/S_+)$ it is necessary to consider the limit $S_+/L \to 0$ which 
is difficult to reach numerically. Approximations thereof are shown in Fig.~\ref{fig:casimir-pp-SnSp}.

This shows how the universal Casimir amplitude can be tuned between its homogeneous value 
$\Delta^{++}_0$ and its maximal value $(\Delta^{++})_{\mathit{max}} \simeq 0$ via patterning. In the 
limit $S_+/L \to 0$ with $S_+/S_- = const$, i.e, for fixed stripe widths and increasing film thickness, 
the periodically structured substrates appear as substrates characterized by laterally averaged, 
effective homogeneous surface fields. For $S_+ = S_-$ the lateral average yields a vanishing effective
surface field, mimicking a Dirichlet (ordinary) boundary condition. For Dirichlet-Dirichlet boundary
conditions the Casimir force vanishes within mean field theory as used here; fluctuations yield a small
attractive force.\cite{Krech1991, Krech1992} This reasoning is consistent with the vanishing of
$\Delta^{++}(S_-/S_+, S_+/L)$ at $S_+/S_- = 1$ for $S_+/L \to 0$ (Fig.~\ref{fig:casimir-pp-SnSp}).
For $S_+ \neq S_-$ the lateral average yields a non-vanishing effective surface field of equal sign 
on both substrates, so that, the Casimir amplitude remains negative (i.e., attractive) for 
$S_+/L \to 0$ even within mean field theory (Fig.~\ref{fig:casimir-pp-SnSp}).

\subsection{\label{sec:pm}Forces for the $+-$~configuration}
In this subsection we study two parallel substrates with their chemical patterns being the opposite 
of each other, i.e., a positive stripe of width $\tilde{S}_+$ opposes a negative stripe of the same 
width (see Fig.~\ref{fig:systems}(b)).

\begin{figure}
   \begin{center}
      \epsfig{file=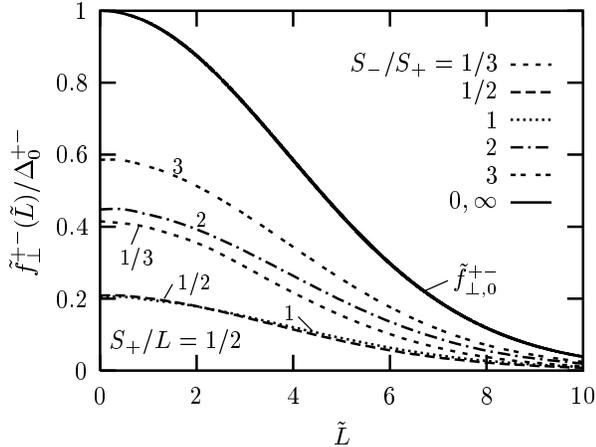,width=0.48\textwidth}
      \caption{\label{fig:pm-SpL1_2-SnSpx}Scaling function $\tilde{f}^{+-}_\perp(\tilde{L})$ of the 
               force between two oppositely patterned substrates (Fig.~\ref{fig:systems}(b)) normalized 
               by $\Delta^{+-}_0$ as a function of the scaled distance $\tilde{L}=L/\xi$ between 
               the substrates for different ratios $S_-/S_+$ of the stripe widths (dashed curves) in 
               comparison with the scaling function $\tilde{f}^{+-}_{\perp,0}(\tilde{L})$ of the force 
               between two homogeneous substrates (solid curve) with a fixed ratio $S_+/L=1/2$.}

   \end{center}
\end{figure}
Figure~\ref{fig:pm-SpL1_2-SnSpx} shows how the scaling function $\tilde{f}^{+-}_\perp(\tilde{L})$ of 
the force in units of $\Delta^{+-}_0$ depends on the ratio $S_-/S_+$ of the stripe widths for a 
fixed ratio $S_+/L$. As expected this force is repulsive. 
As in the case of the $++$~configuration the force $\tilde{f}^{+-}_\perp(\tilde{L})$ is reduced 
compared to the force $\tilde{f}^{+-}_{\perp,0}(\tilde{L})$ between two homogeneous substrates,
takes a finite value in the limit $\tilde{L} \to 0$, and vanishes exponentially for $\tilde{L} \to \infty$.
\begin{figure}
   \begin{minipage}{0.48\textwidth}
   \begin{center}
      \epsfig{file=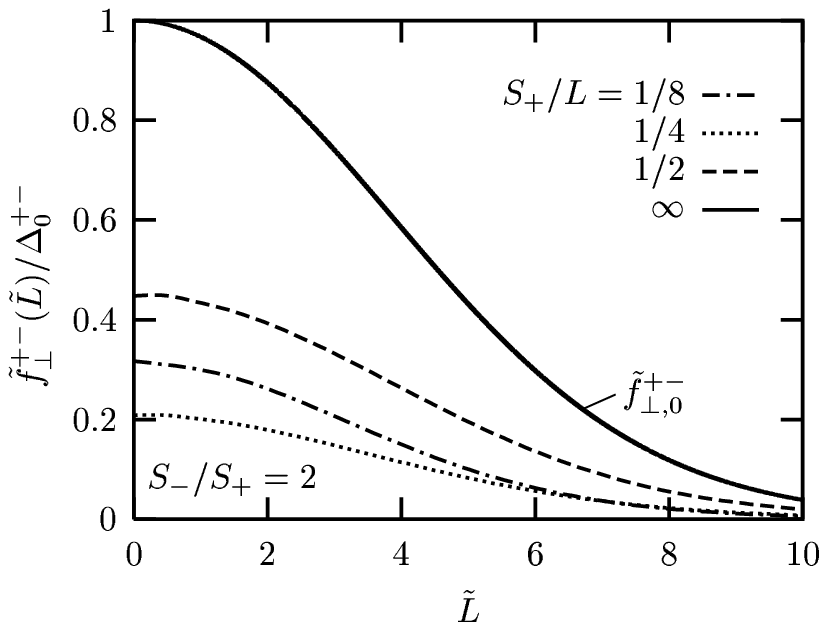,width=\textwidth}
      \caption{\label{fig:pm-SpLx-SnSp2_1}Scaling function $\tilde{f}^{+-}_\perp(\tilde{L})$ of the 
               force between two oppositely patterned substrates (Fig.~\ref{fig:systems}(b)) normalized 
               by $\Delta^{+-}_0$ for different ratios $S_+/L$ and a fixed ratio $S_-/S_+ = 2$.}
      \epsfig{file=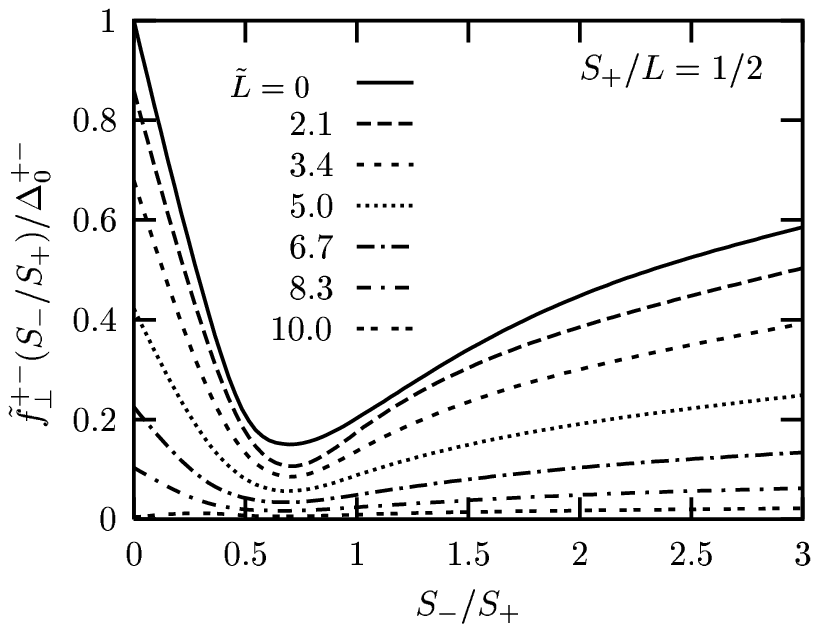,width=\textwidth}
      \caption{\label{fig:pm-SnSp-L}Scaling function $\tilde{f}^{+-}_\perp(S_-/S_+)$ of the force 
               between two oppositely patterned substrates normalized by $\Delta^{+-}_0$ as a
               function of the stripe width ratio $S_-/S_+$ for various fixed scaled distances 
               $\tilde{L}= L/\xi$ between the substrates and for a ratio $S_+/L = 1/2$. The dependence 
               of the force on the ratio $S_-/S_+$ becomes weaker with increasing distance.}
   \end{center}
   \end{minipage}
\end{figure}
For $S_-/S_+ \to \infty$ the force tends to the force $\tilde{f}^{+-}_{\perp,0}(\tilde{L})$ between 
two homogeneous, antagonistic substrates. For small ratios $S_-/S_+ \to 0$ the negative stripes vanish 
as the ratio $S_+/L$ is fixed and therefore also in this limit the homogeneous case is reached.

When comparing the forces for systems with the same ratio $S_-/S_+$ of the stripe widths but with 
different ratios $S_+/L$ (see Fig.~\ref{fig:pm-SpLx-SnSp2_1}) (i.e., systems with different values of 
scaled widths $\tilde{S}_+$ and $\tilde{S}_-$ of the stripes) the same behavior as for the 
$++$~configuration is exhibited: for fixed distances $\tilde{L} \gtrsim 7$ and increasing stripe widths 
the force acting in the system tends monotonicly towards the homogeneous regime, for smaller distances 
$\tilde{L}$ the trend is partly inverted due to competing effects caused by reduced healing for very 
small stripes.

\begin{figure}
   \begin{center}
      \epsfig{file=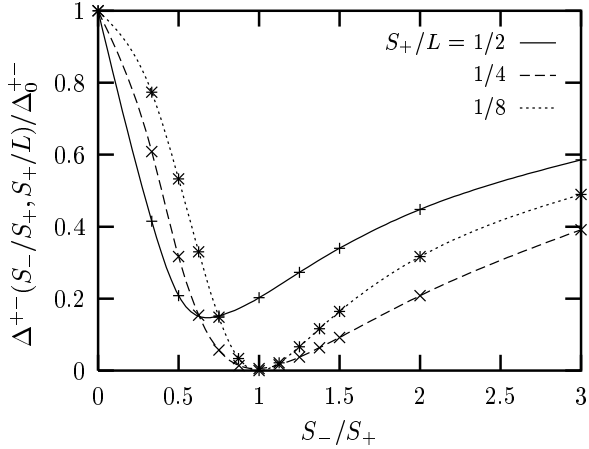,width=0.48\textwidth}
      \caption{\label{fig:casimir-pm-SnSp}Generalized Casimir amplitude $\Delta^{+-}(S_-/S_+, S_+/L)$ 
               for a critical fluid confined between two oppositely patterned substrate normalized by
               $\Delta^{+-}_0$ as a function of the stripe width ratio $S_-/S_+$ for various ratios 
               $S_+/L$. The curve for $S_+/L = 1/8$ serves as an approximation for the actual generalized
               Casimir amplitude $\Delta^{+-}(S_-/S_+)$ corresponding to the limit $S_+/L \to 0$. 
               $\Delta^{+-}(S_-/S_+)$ varies between $\Delta^{+-}_0>0$ for $S_-/S_+ =0, \infty$ and its
               minimum value $(\Delta^{+-})_{\mathit{min}} \simeq 0$ (see main text).}
   \end{center}
\end{figure}
Figure~\ref{fig:pm-SnSp-L} shows the dependence on the stripe width ratio $S_-/S_+$. It becomes 
weaker with increasing scaled distance $\tilde{L}$ between the substrates. The limits $S_-/S_+ \to 0$ 
and $S_-/S_+ \to \infty$ correspond to the force $\tilde{f}^{+-}_{\perp,0}(\tilde{L})$ between two 
homogeneous substrates. For $\tilde{L}=0$ Fig.~\ref{fig:pm-SnSp-L} leads to a universal scaling 
function $\Delta^{++}(S_-/S_+ ,S_+/L)$ which in the limit of large $L$, $S_-/S_+ \to 0$, and 
$S_+/L = const$ or in the limit of large $L$, $S_-/S_+ \to \infty$, and $S_+/L = const$ reduces to 
the Casimir amplitude $\Delta^{+-}_0$ for homogeneous antagonistic systems. A generalized Casimir 
amplitude $\Delta^{+-}(S_-/S_+)$ as the analogue of $\Delta^{++}(S_-/S_+)$ (see 
Fig.~\ref{fig:casimir-pp-SnSp}) is obtained in the limit $S_+/L \to 0$. A numerical approximation thereof 
is shown in Fig.~\ref{fig:casimir-pm-SnSp}.
Like for the $++$~configuration (Fig.~\ref{fig:casimir-pp-SnSp}) the actual generalized Casimir amplitude 
$\Delta^{+-}(S_-/S_+)$ for the $+-$~configuration vanishes for a stripe width ratio $S_+/S_- = 1$ in 
the limit $S_+/L \to 0$ corresponding effectively to Dirichlet boundary conditions evoking a vanishing 
force within mean field theory (Fig.~\ref{fig:casimir-pm-SnSp}).

\subsection{\label{sec:ph}Forces for the $ph$~configuration}
The configuration of a patterned substrate at $z=0$ parallel to a homogeneous substrate with a positive 
surface field $h_1 = + \infty$ positioned at $z=L$ is of special interest because it can be realized by 
a binary liquid mixture with its bulk in the vapor phase forming a wetting film near demixing on a 
patterned substrate at $z=0$ with film thickness $L$ such that at the free liquid-vapor interface the 
vapor plays the role of the homogeneous substrate. This kind of experiment is a natural extension of 
the critical Casimir experiments as carried out in Ref.~\onlinecite{Pershan2005} on a homogeneous 
substrate.

\begin{figure}
   \begin{minipage}{0.48\textwidth}
   \begin{center}
      \epsfig{file=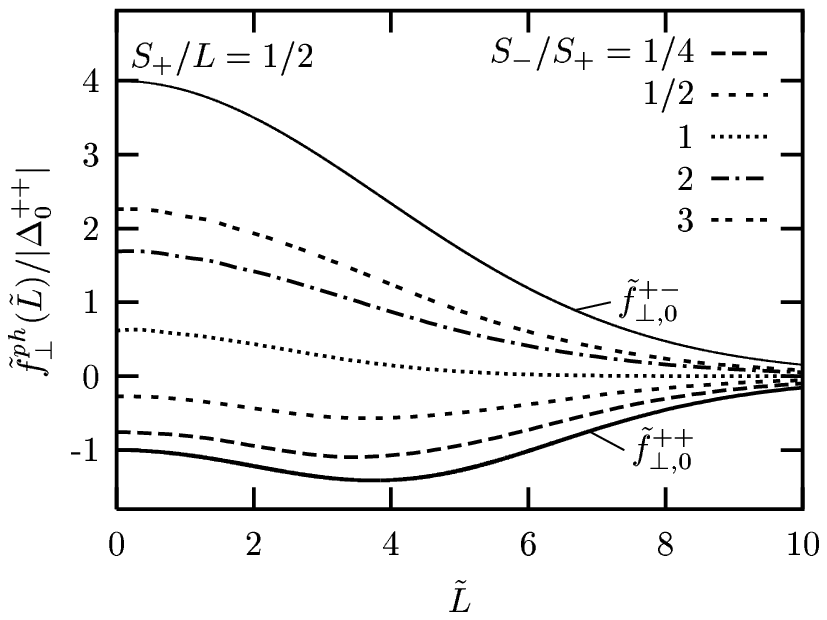,width=\textwidth}
      \caption{\label{fig:ph-SpL1_2-SnSpx}Scaling function $\tilde{f}^{ph}_\perp(\tilde{L})$ of the force
               between a patterned and a homogeneous substrate with a positive surface field normalized by
               $|\Delta^{++}_0|$ as a function of the scaled distance between the substrates 
               for different stripe width ratios $S_-/S_+$ and a fixed ratio 
               $S_+/L=1/2$.}
      \epsfig{file=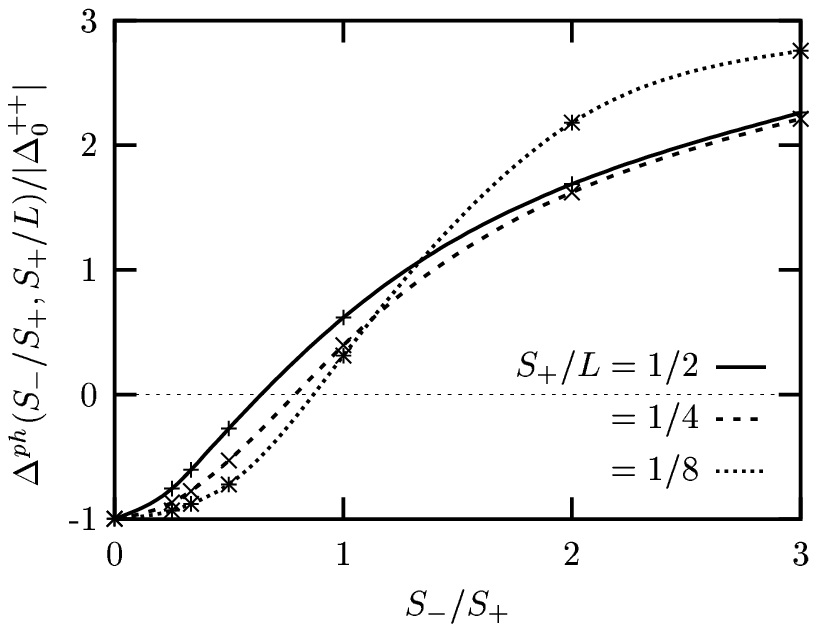,width=\textwidth}
      \caption{\label{fig:casimir-ph-SnSp}Generalized Casimir amplitude $\Delta^{ph}(S_-/S_+, S_+/L)$ 
               for a critical fluid confined between a patterned and a homogeneous substrate normalized 
               by $|\Delta^{++}_0|$ as a function of the stripe width ratio $S_-/S_+$ for various
               ratios $S_+/L$. As in Fig.~\ref{fig:casimir-pp-SnSp} and \ref{fig:casimir-pm-SnSp} the 
               curve for $S_+/L = 1/8$ serves as an approximation for the actual generalized
               Casimir amplitude $\Delta^{ph}(S_-/S_+)$ corresponding to the limit $S_+/L \to 0$. 
               $\Delta^{ph}(S_-/S_+)$ varies between $\Delta^{++}_0<0$ for $S_-/S_+ =0$ and 
               $\Delta^{+-}_0>0$ for $S_-/S_+ =\infty$. Consequently $\Delta^{ph}$ has a zero and we 
               find $\Delta^{ph}(S_-/S_+)=0$ for $S_- \simeq S_+$.}
   \end{center}
   \end{minipage}
\end{figure}
Figure~\ref{fig:ph-SpL1_2-SnSpx} illustrates the dependence of the scaling function 
$\tilde{f}^{ph}_\perp(\tilde{L})$ of the force between a patterned and a homogeneous substrate on 
the stripe widths ratio $S_-/S_+$ for a fixed ratio $S_+/L$. For increasing negative stripe widths 
and fixed separations of the plates (i.e., $S_-/S_+ \to \infty$) the force becomes more repulsive and 
reaches the limiting case of a system with two homogeneous substrates with antiparallel surface fields. 
In contrast, for vanishing negative stripes (i.e., $S_-/S_+ \to 0$) the force is attractive and converges 
to the force between two homogeneous substrates with parallel surface fields.

At the critical point, i.e., for $\tilde{L}=0$, and for large $L$ with $S_+/L = const$ 
the scaling function $\tilde{f}^{ph}_\perp$ reduces to a universal function $\Delta^{ph}(S_-/S_+, S_+/L)$ 
(see Fig.~\ref{fig:casimir-ph-SnSp}) which interpolates between $\Delta^{++}_0$ at $S_-/S_+ = 0$ and 
$\Delta^{+-}_0$ at $S_-/S_+ = \infty$ and therefore has a zero. For $S_+/L \to 0$, according to 
Fig.~\ref{fig:casimir-ph-SnSp} $\Delta^{ph}(S_-/S_+)$ vanishes for $S_- / S_+ \simeq 1$. For this 
stripe width ratio, at the critical point the Casimir force  decays faster than $L^{-d}$. 

\begin{figure}
   \begin{minipage}{0.48\textwidth}
   \begin{center}
      \epsfig{file=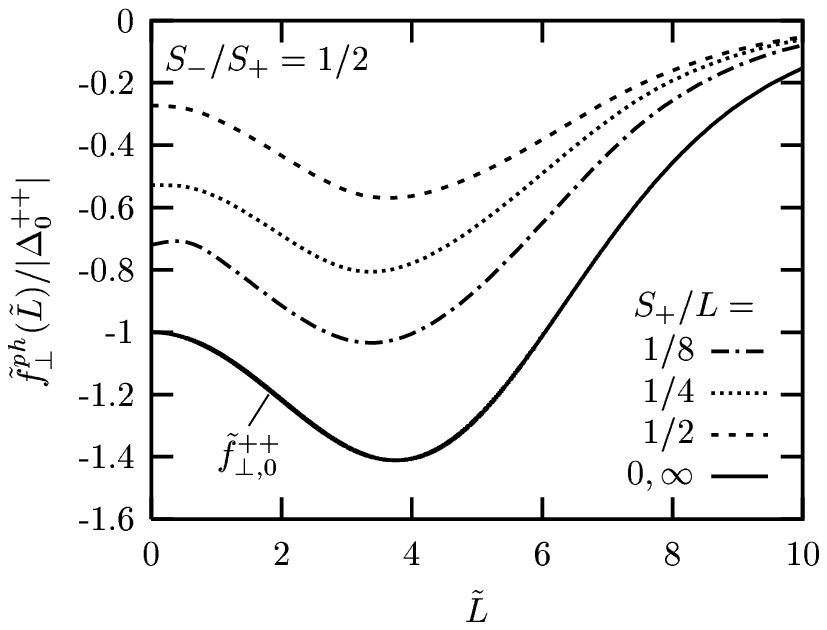,width=\textwidth}
      \caption{\label{fig:ph-SpLx-SnSp1_2}Scaling function $\tilde{f}^{ph}_\perp(\tilde{L})$ of the 
               force between a patterned and a homogeneous substrate with $h_1=+\infty$ normalized by
               $|\Delta^{++}_0|$ for different ratios $S_+/L$ with $S_-/S_+ = 1/2$ fixed.}
      \epsfig{file=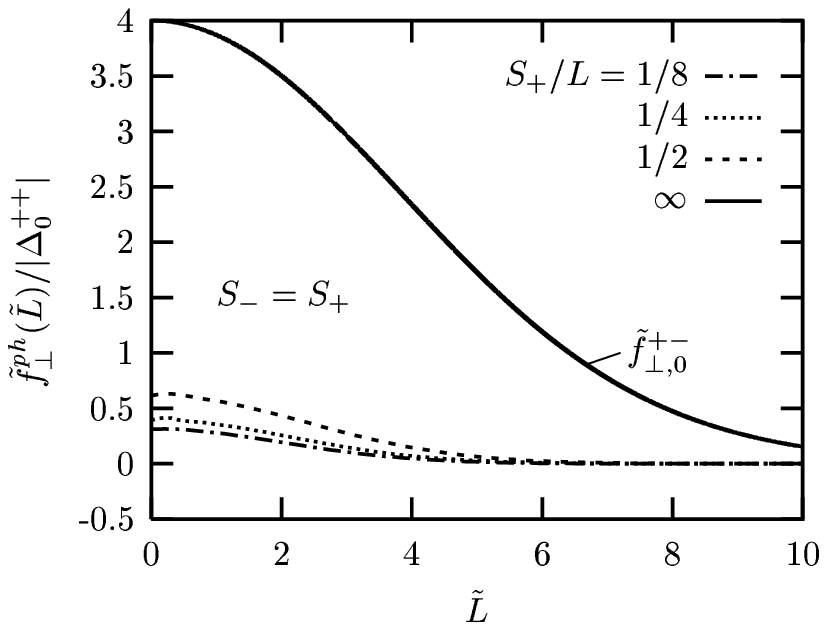,width=\textwidth}
      \caption{\label{fig:ph-SpLx-SnSp1_1}Scaling function $\tilde{f}^{ph}_\perp(\tilde{L})$ of the
               force between a patterned and a homogeneous substrate with $h_1=+\infty$ normalized by 
               $|\Delta^{++}_0|$ for different ratios $S_+/L$ with $S_- = S_+$ fixed.}
      \epsfig{file=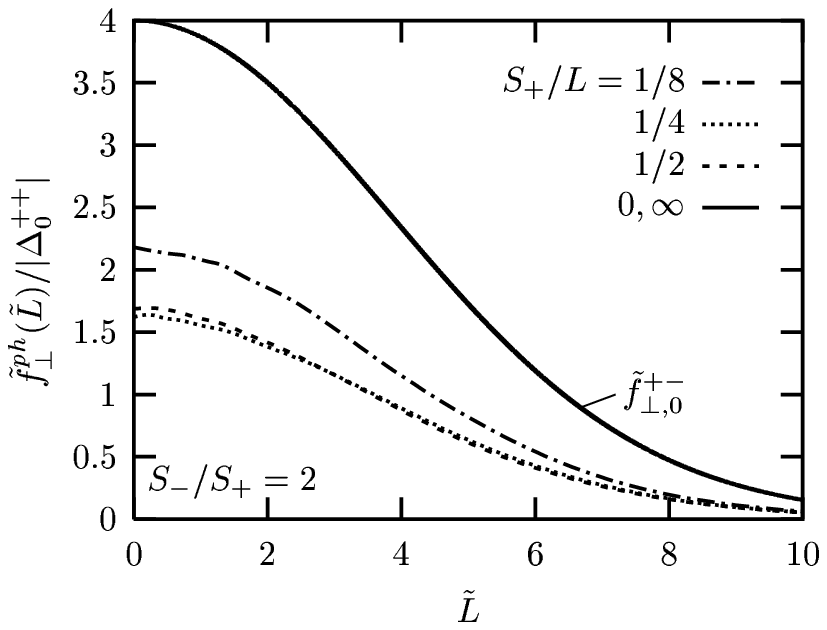,width=\textwidth}
      \caption{\label{fig:ph-SpLx-SnSp2_1}Scaling function $\tilde{f}^{ph}_\perp(\tilde{L})$ of the
               force between a patterned and a homogeneous substrate  with $h_1=+\infty$ normalized by 
               $|\Delta^{++}_0|$ for different ratios $S_+/L$ with $S_-/S_+ = 2$ fixed.}
   \end{center}
   \end{minipage}
\end{figure}

The dependence of the force on the scaled distance between the substrates is shown in 
Figs.~\ref{fig:ph-SpLx-SnSp1_2}, \ref{fig:ph-SpLx-SnSp1_1}, and \ref{fig:ph-SpLx-SnSp2_1} for three 
different ratios $S_+/L$ and three different ratios $S_-/S_+$. For simultaneously decreasing positive 
and negative stripe widths $\tilde{S}_+ \to 0$ and $\tilde{S}_- \to 0$ the force tends - depending on 
the ratio $S_-/S_+$ - towards those for the different homogeneous cases, i.e., for $S_-/S_+ > 1$ 
towards the force between plates with antiparallel surface fields and for $S_-/S_+ < 1$ towards the 
force between plates with parallel surface fields. The larger the ratio $S_+/L$, the larger is the 
scaled stripe width $\tilde{S}_+$ for $\tilde{L}$ fixed and the more developed is the healing of the 
order parameter distribution across the chemical steps. Therefore the absolute value of the force is 
reduced as compared to smaller stripe widths.

\subsection{\label{sec:compare}Comparison of the configurations}
\begin{widetext}
\begin{figure}
   \begin{minipage}{\textwidth}
   \begin{center}
      \epsfig{file=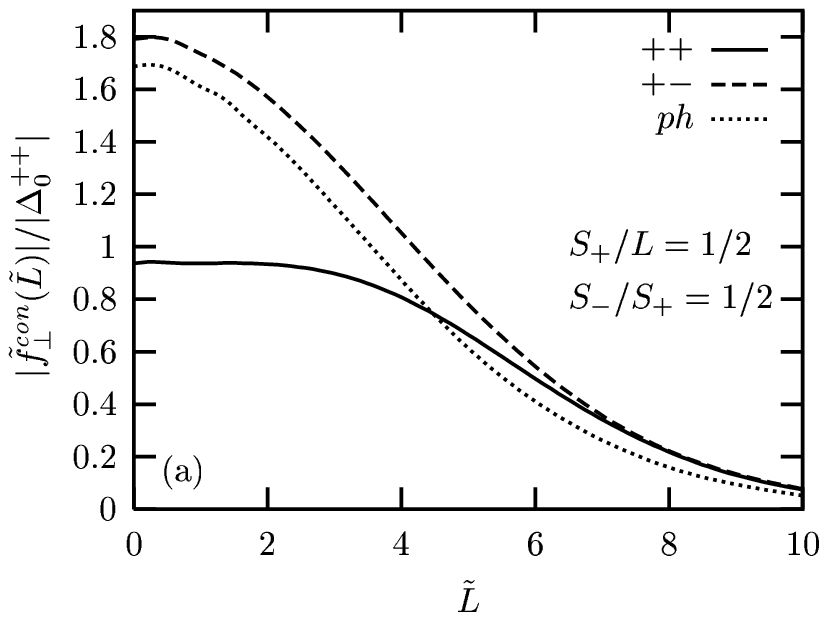,width=0.48\textwidth}
      \epsfig{file=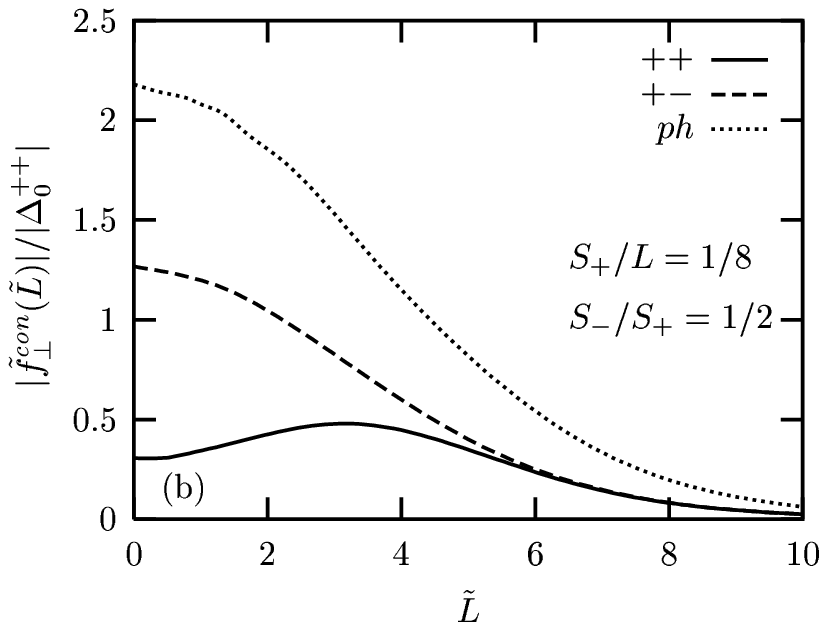,width=0.48\textwidth}
      \epsfig{file=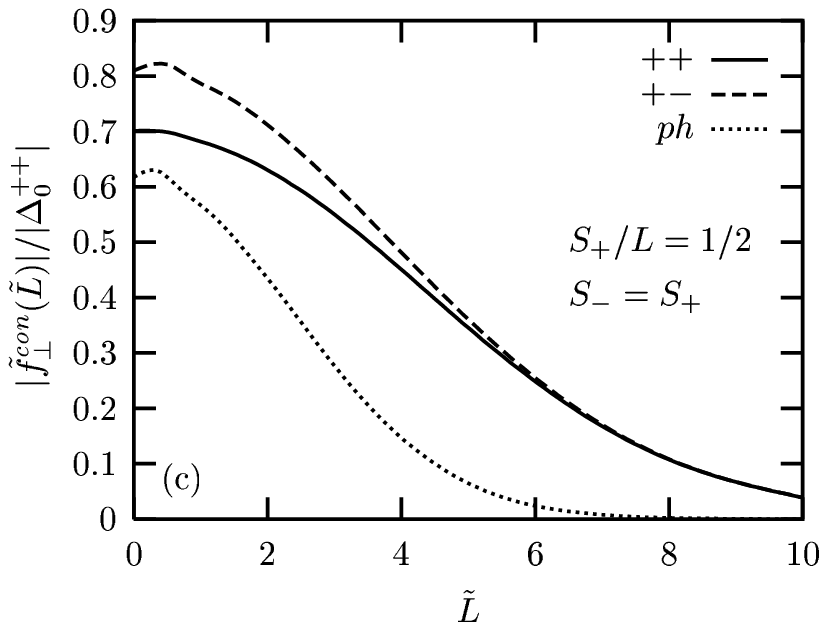,width=0.48\textwidth}
      \epsfig{file=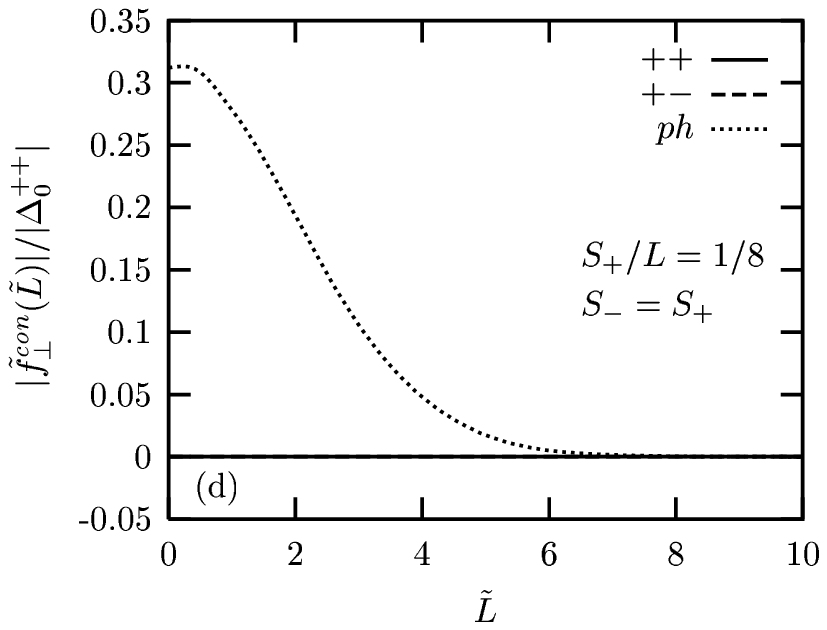,width=0.48\textwidth}
      \caption{\label{fig:vgl-SpLx-SnSpy}Comparison of the absolute values of the scaling functions
               $\tilde{f}^{con}_\perp(\tilde{L})$ for the forces between patterned substrates (where 
               $con=++,+-,ph$) normalized by $|\Delta^{++}_0|$, (a) for the ratios $S_+/L= 1/2$ and 
               $S_-/S_+ = 2$, (b) $S_+/L= 1/8$ and $S_-/S_+ = 2$, (c) $S_+/L= 1/2$ and 
               $\tilde{S}_- = \tilde{S}_+$, and (d) $S_+/L= 1/8$ and $\tilde{S}_- = \tilde{S}_+$; on this 
               scale the values for both the $++$ and the $+-$~configuration cannot be distinguished from 
               zero.}
   \end{center}
   \end{minipage}
\end{figure}
\end{widetext}

\begin{figure}
   \begin{center}
      \epsfig{file=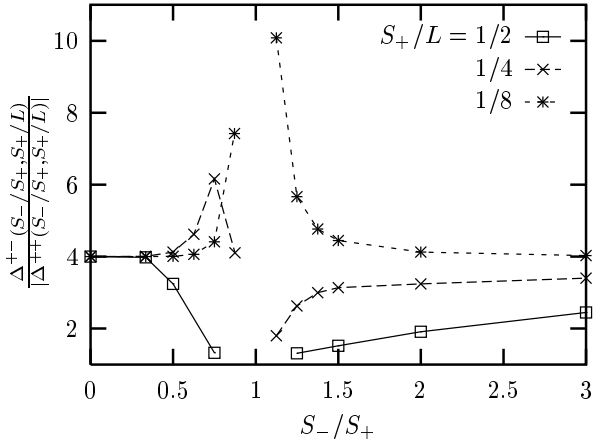,width=0.48\textwidth}
      \caption{\label{fig:compare-amplitudes}Ratio 
               $\Delta^{+-}(S_-/S_+, S_+/L) / |\Delta^{++}(S_-/S_+, S_+/L)|$ of the generalized Casimir
               amplitudes for the $+-$~configuration and the $++$~configuration as a function of the 
               ratio $S_-/S_+$ of the stripe widths. The limits $S_-/S_+ = 0, \infty$ correspond to 
               homogeneous substrates for which the ratio equals $\Delta^{+-}_0 / |\Delta^{++}_0| = 4$ 
               (see Subsec.~\ref{subsec:force-stresstensor}). It is difficult to evaluate numerically 
               this ratio for $S_-/S_+ \simeq 1$ because there both $\Delta^{+-}(S_-/S_+, S_+/L)$ and
               $\Delta^{++}(S_-/S_+, S_+/L)$ are very small.}
   \end{center}
\end{figure}

In this subsection we compare the forces for the aforementioned configurations $++$, $+-$ and $ph$.
First, we find that the forces for the $++$~configuration and the $+-$~configuration are of the same 
order of magnitude. For the same stripe geometries the absolute value of the force scaling function
$\tilde{f}^{+-}_\perp(\tilde{L})$ for the $+-$~configuration is always larger than the absolute 
value of the force scaling function $\tilde{f}^{++}_\perp(\tilde{L})$ for the $++$~configuration (see  
Fig.~\ref{fig:vgl-SpLx-SnSpy}).
The ratio $\Delta^{+-}(S_-/S_+, S_+/L)/|\Delta^{++}(S_-/S_+, S_+/L)|$ of the generalized Casimir 
amplitudes for the $++$~configuration and the $+-$~configuration is smaller than the factor $4$ found 
for the forces between homogeneous substrates (for details see Fig.~\ref{fig:compare-amplitudes}). 
This is caused by the healing effect for the order parameter distribution.

For a ratio $S_+/L = 1/2$ 
(see Fig.~\ref{fig:vgl-SpLx-SnSpy}(a)) and a ratio $S_-/S_+ < 1$ the force 
$\tilde{f}^{ph}_\perp(\tilde{L})$ between the substrates for the $ph$~configuration is larger than 
the force $\tilde{f}^{++}_\perp(\tilde{L})$ for the $++$~configuration but smaller than the force 
$\tilde{f}^{+-}_\perp(\tilde{L})$ for the $+-$~configuration. If the stripes become smaller 
(i.e., $S_+/L \to 0$) 
(see Fig.~\ref{fig:vgl-SpLx-SnSpy}(b)) the force between the substrates for the $ph$~configuration 
becomes larger than for the $+-$ and the $+-$~configuration. 

For the ratio $S_-/S_+ = 1$ these effects are even more pronounced due to the antisymmetry of the 
order parameter profiles:
Fig.~\ref{fig:vgl-SpLx-SnSpy}(c) shows the scaling function in the case \mbox{$S_+/L = 1/2$} for which 
the force between the substrates for the $ph$~configuration is even smaller than both the force for the 
$++$ and the force for the $+-$~configuration. With decreasing stripe widths (see 
Fig.~\ref{fig:vgl-SpLx-SnSpy}(d)) the force between the substrates for the $ph$~configuration 
is significantly larger than for the $++$ and the $+-$~configuration. For these very small stripes and
for $S_-/S_+ = 1$ the structured substrate effectively appears as a surface with a surface field tending 
to zero which is effectively equivalent to the Dirichlet boundary condition. In contrast the homogeneous 
substrate in the $ph$~configuration still has a positive surface field such that for the 
$ph$~configuration a force remains even in the limit $S_+/L \to 0$. 

The Casimir amplitude $\Delta^{o+}$ of the force between a substrate with ordinary (Dirichlet) boundary 
condition and a substrate with $+$~boundary condition within mean field theory is given by \cite{Krech1997}
\begin{equation}
   \Delta^{o+} = -\frac{1}{4} \Delta^{++}_0 \,.
\end{equation}
The value $\tilde{f}^{ph}_\perp(0) / |\Delta^{++}_0|$ for $S_+/L = 1/8$ in 
Fig.~\ref{fig:vgl-SpLx-SnSpy}(d) is still above this value because the limit $S_+/L \to 0$ is not yet 
reached.

Figure~\ref{fig:compare-amplitudes} shows the dependence of the ratio 
$\Delta^{+-}(S_-/S_+, S_+/L) / |\Delta^{++}(S_-/S_+, S_+/L)|$ of the generalized Casimir amplitudes 
for the $+-$~configuration and the $++$~configuration on the stripe width ratio $S_-/S_+$.
For $S_-/S_+=0, \infty$ the substrates are homogeneous and therefore in these limits 
$\Delta^{+-}(S_-/S_+, S_+/L) / |\Delta^{++}(S_-/S_+, S_+/L)| = \Delta^{+-}_0 / |\Delta^{++}_0| = 4$. 
(Since both $\Delta^{++}$ and $\Delta^{+-}$ are very small for $S_-/S_+ \simeq 1$ (see 
Figs.~\ref{fig:casimir-pp-SnSp} and \ref{fig:casimir-pm-SnSp}), the numerical estimates for this 
ratio become unreliable there.)

\subsection{\label{sec:ps}Lateral shift}
Finally we consider two parallel substrates endowed with the same pattern but laterally shifted relative 
to each other by a distance $D$ in $x$~direction, $\tilde{D}= D/ \xi$ (see Fig.~\ref{fig:systems}(d)). 
In order to limit the number of relevant length scales here we choose $S_+ = S_- \equiv S$ and 
$\tilde{S}_+ = \tilde{S}_- \equiv \tilde{S}$, respectively, so that with the relative shift 
\mbox{$\delta \equiv D / S = \tilde{D} / \tilde{S}$} the set of variables is $(\tilde{L}, \tilde{S}, \tilde{D})$ 
or $(\tilde{L}, \tilde{S}, \delta)$. This set of variables is chosen to simplify the presentation
of the scaling functions of the force but it has the disadvantage that upon approaching the critical 
point both the scaled distance $\tilde{L}$ and the scaled stripe width $\tilde{S}$ vanish which makes it 
difficult to calculate the forces between the substrates at the critical point. 
In order to obtain the forces at $T_c$ the set of variables 
$(\tilde{L}, \tilde{S}/\tilde{L}, \tilde{D} / \tilde{S}) = (\tilde{L}, S/L, \delta)$ in which only the 
scaled distance $\tilde{L}$ vanishes in the limit $T \to T_c$ is chosen.
For this configuration we study the force $\tilde{f}^{ps}_\perp(\tilde{L}, \tilde{S}, \delta)$ normal to the 
substrate like for the previous configurations. But in this case in addition a lateral force 
$\tilde{f}^{ps}_\parallel(\tilde{L}, \tilde{S}, \delta)$ parallel to the substrates emerges.

\subsubsection{Normal force}
\begin{figure}
   \begin{minipage}{0.48\textwidth}
   \begin{center}
      \epsfig{file=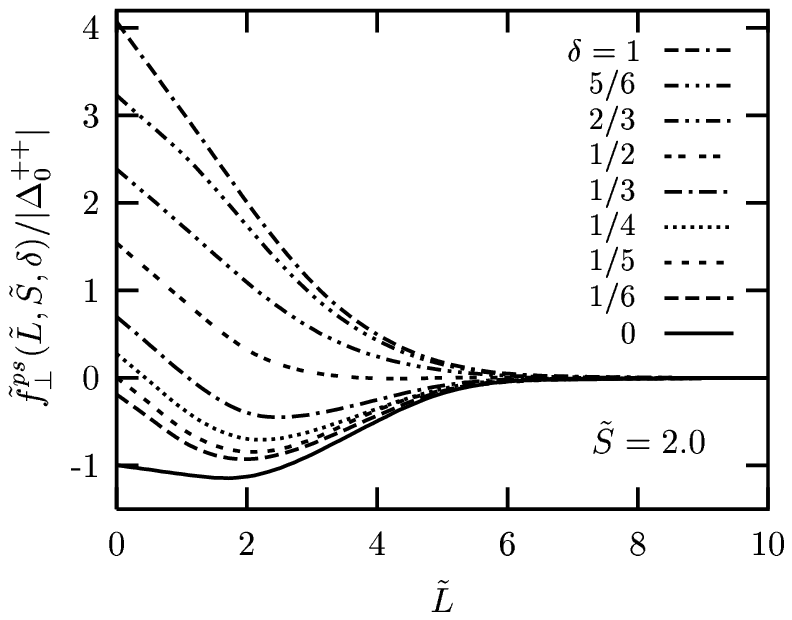,width=\textwidth}
      \caption{\label{fig:ps-SD_L}Scaling function $\tilde{f}^{ps}_\perp(\tilde{L}, \tilde{S}, \delta)$ 
               of the normal force between two substrates patterned identically with positive and 
               negative stripes of widths $S_+ = S_- = S$ and $\tilde{S}=2.0$ for various lateral shifts
               $\delta = D/S = \tilde{D} / \tilde{S}$. The scaling function of the force is normalized by 
               $|\Delta^{++}_0|$. The values for $\tilde{L}=0$ are obtained by taking the values for small
               $\tilde{L}$ and extrapolating them linearly to $\tilde{L}=0$.}
      \epsfig{file=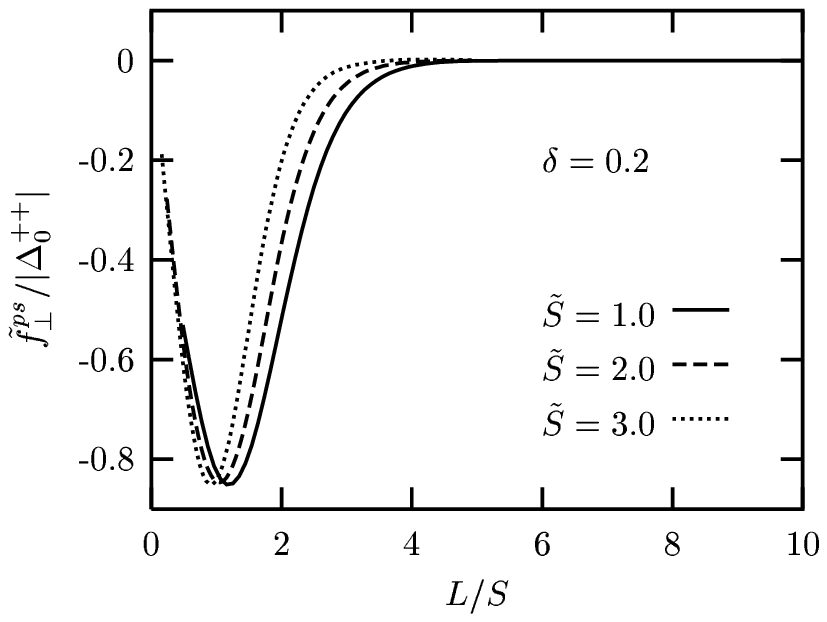,width=\textwidth}
      \caption{\label{fig:ps-S1S2D_L-scaled}Scaling function $\tilde{f}^{ps}_\perp$ of the 
               normal force between two substrates patterned with positive and negative stripes of equal
               width $\tilde{S}=1.0$ (solid line), $\tilde{S}=2.0$ (dashed line), and $\tilde{S}=3.0$ 
               (dotted line) as function of $L/S$ and with a relative shift of $\delta = 0.2$. 
               Here $\tilde{f}^{ps}_\perp$ is considered as a scaling function of the scaling variables
               $L/S$, $\tilde{S}$, and $\delta$. The scaling function of the force is normalized by
               $|\Delta^{++}_0|$.}
   \end{center}
   \end{minipage}
\end{figure}
Figure~\ref{fig:ps-SD_L} illustrates the influence of the relative shift on the normal force 
$\tilde{f}^{ps}_\perp(\tilde{L}, \tilde{S}, \delta)$. For a vanishing shift $\delta = 0$ the system 
is in the patterned $++$~configuration and the force is attractive for all scaled distances $\tilde{L}$ 
between the substrates. With increasing shift $\delta$ (i.e., $D \to S$) the force increases and becomes 
repulsive for all scaled distances $\tilde{L}$ until the corresponding patterned $+-$~configuration
is reached for $\delta=1$. For $\delta \gtrsim 0.4$ the force becomes repulsive for all $\tilde{L}$.

The influence of the scaled stripe width $\tilde{S}$ on the scaling function 
$\tilde{f}^{ps}_\perp(\tilde{L}, \tilde{S}, \delta)$ is shown in Fig.~\ref{fig:ps-S1S2D_L-scaled} 
through the dependence of  $\tilde{f}^{ps}_\perp$ on the ratio $L/S$. The curves for the scaled stripe 
widths $\tilde{S}=1.0$, $\tilde{S}=2.0$, and $\tilde{S}=3.0$ with the same relative shift $\delta = 0.2$ 
do not coincide. This demonstrates that the force $\tilde{f}^{ps}_\perp(\tilde{L}, \tilde{S}, \delta)$ 
does not reduce to an effective scaling function of only the two variables $L/S$ and $\delta$.

\begin{figure}
   \begin{center}      
      \epsfig{file=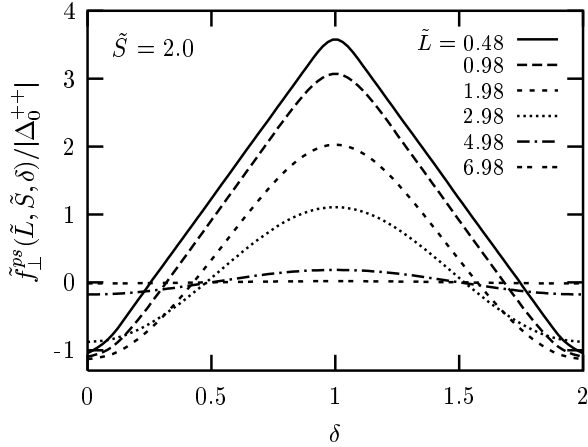,width=0.48\textwidth}
      \caption{\label{fig:ps-SL_D}Same data as in Fig.~\ref{fig:ps-SD_L} shown as function of the relative
               shift $\delta = D/S$.}
   \end{center}
\end{figure}

\begin{figure}
      \epsfig{file=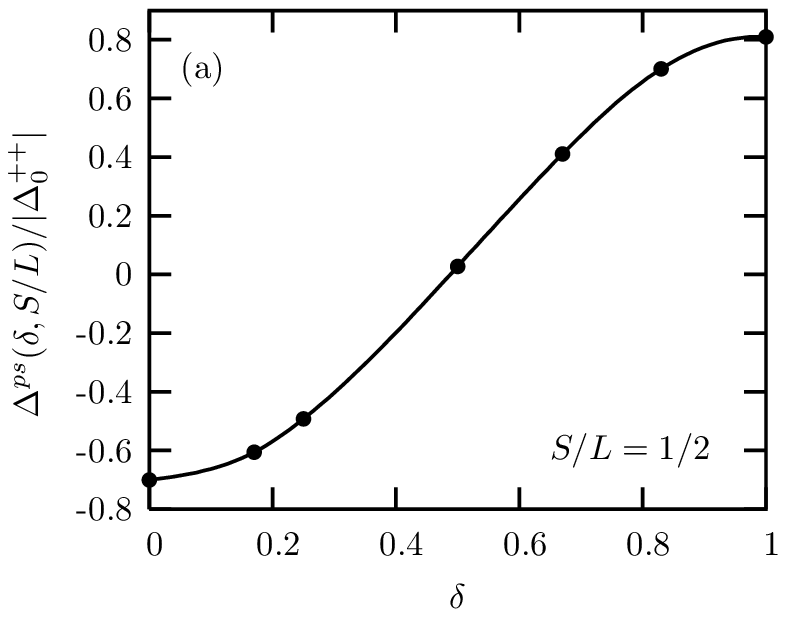,width=0.48\textwidth} \\[-5mm]
      \hspace*{-4mm}\epsfig{file=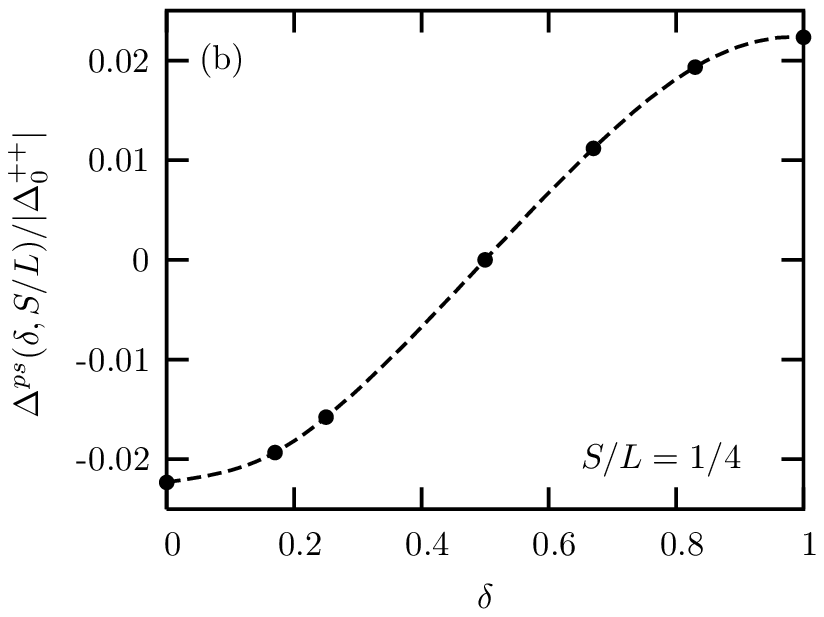,scale=0.88} \\[-5mm]
      \hspace*{-8mm}\epsfig{file=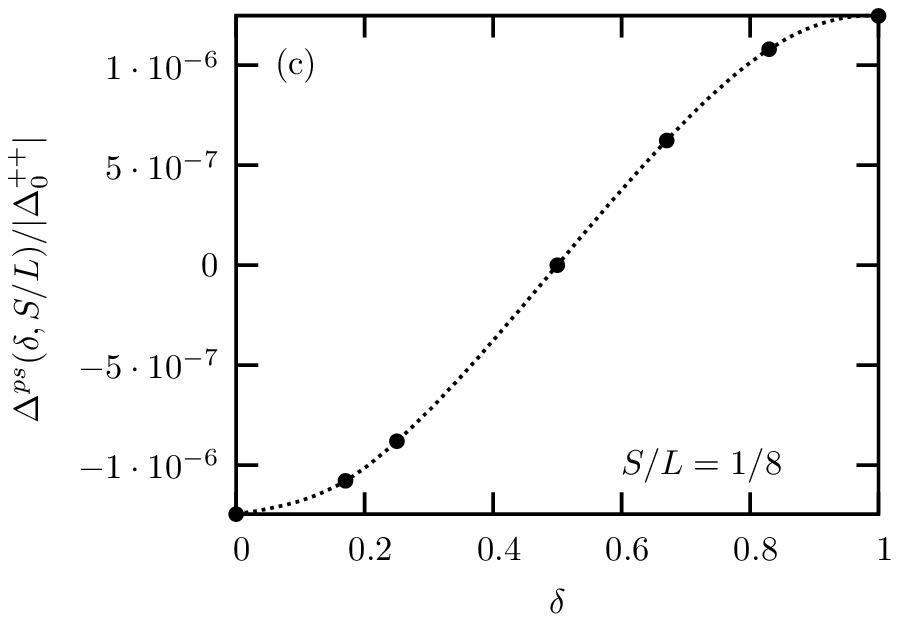,scale=0.88} \\[-7mm]
      \caption{\label{fig:casimir-ps-SL1_x}Generalized Casimir amplitude $\Delta^{ps}(\delta, S/L)$ for a 
               critical fluid confined between two identically structured substrates with relative shift 
               $\delta$ normalized by $|\Delta^{++}_0|$ as a function of the relative shift $\delta$ 
               (a) for a ratio $S/L=1/2$, (b) for a ratio $S/L=1/4$, (c) for a ratio $S/L=1/8$; 
               $S_+ =S_- \equiv S$.}
\end{figure}

In Fig.~\ref{fig:ps-SL_D} the data contained in Fig.~\ref{fig:ps-SD_L} are shown as to visualize 
the dependence on the relative shift $\delta = D/S$. For small scaled distances $\tilde{L}$ the force
scaling function $\tilde{f}^{ps}_\perp$ depends strongly on the relative shift, with the curves 
interpolating between the value of the force scaling function $\tilde{f}^{++}_\perp(\tilde{L})$ for 
the patterned $++$~configuration ($\delta=0$) and of the force scaling function 
$\tilde{f}^{+-}_\perp(\tilde{L})$ for the patterned $+-$~configuration ($\delta=1$). The dependence 
on $\delta$ disappears with increasing scaled distance $\tilde{L} \to \infty$ between the substrates. 

The normal force at the critical point is characterized by a generalized Casimir amplitude 
$\Delta^{ps}(\delta, S/L)$ shown in Fig.~\ref{fig:casimir-ps-SL1_x} for the ratios $S/L=1/2, 1/4, 1/8$. 
The generalized Casimir amplitude $\Delta^{ps}(\delta, S/L)$ varies as a function of the relative 
shift $\delta$ between \mbox{$\Delta^{++}(S_-/S_+ =1, S_+/L)$} for $\delta=0$ and 
$\Delta^{+-}(S_-/S_+ =1, S_+/L)$ for $\delta=1$. Like the generalized Casimir amplitudes 
$\Delta^{++}(S_-/S_+, S_+/L)$ and $\Delta^{+-}(S_-/S_+, S_+/L)$ vanishing for a stripe width ratio 
$S_-/S_+=1$ (Figs.~\ref{fig:casimir-pp-SnSp} and \ref{fig:casimir-pm-SnSp}) in the limit $S_+/L \to 0$ -- 
resembling effective Dirichlet boundary conditions -- , the generalized Casimir amplitude 
$\Delta^{ps}(\delta, S/L)$ vanishes, within mean field theory, in the limit $S/L \to 0$, too, for all 
values of the shift $\delta$ (recall that here $S_+ =S_- \equiv S$).

\subsubsection{Lateral force}
The scaling function $\tilde{f}^{ps}_\parallel(\tilde{L}, \tilde{S}, \delta)$ of the lateral forces 
is calculated according to Eqs.~(\ref{eq:f_xz}) and (\ref{eq:f_vw-P}).
\begin{figure}
   \begin{center}
      \epsfig{file=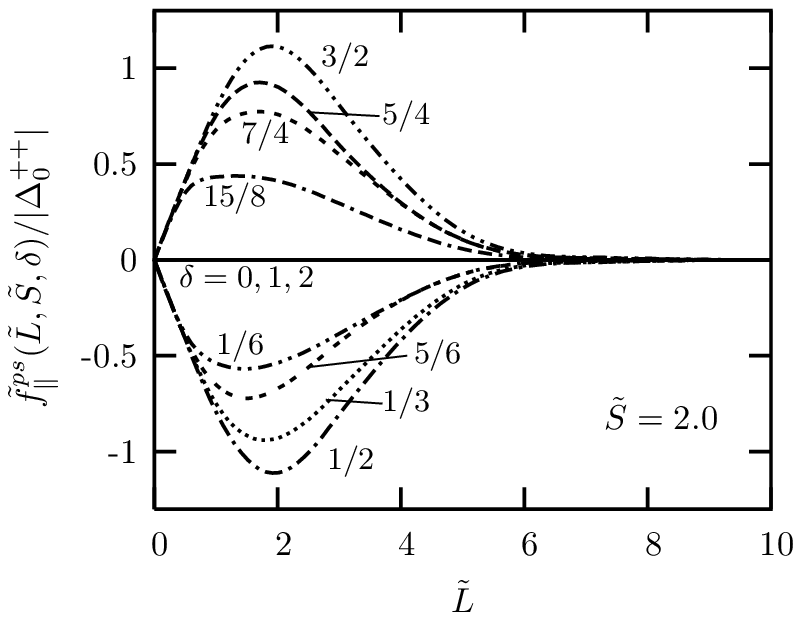,width=0.48\textwidth}
      \caption{\label{fig:ps-tang-SD-S2.0}Scaling function 
               $\tilde{f}^{ps}_\parallel(\tilde{L}, \tilde{S},\delta)$ of the lateral force, normalized
               by $|\Delta^{++}_0|$, between two identically patterned substrates as function of the 
               distance $\tilde{L}$ between the substrates for various relative shifts $\delta$ and
               for a scaled stripe width $\tilde{S}=2.0$.}
   \end{center}
\end{figure}
Figure \ref{fig:ps-tang-SD-S2.0} shows the dependence of the scaling function 
$\tilde{f}^{ps}_\parallel(\tilde{L}, \tilde{S}, \delta)$ for the lateral force on the scaled distance 
$\tilde{L}$ between the substrates for a fixed scaled stripe width $\tilde{S}=2.0$. For the relative shifts 
$\delta=0,2$ and $\delta=1$, which correspond to the patterned $++$ and the patterned $+-$~configuration, 
respectively, the lateral force $\tilde{f}^{ps}_\parallel(\tilde{L}, \tilde{S}, \delta)$ vanishes 
identically. With increasing shifts up to $\delta \approx 0.5$ the force first becomes increasingly 
restoring and for shifts $\delta \gtrsim 0.5$ it decreases again towards 0. For shifts $0 < \delta < 1$ 
the direction of the force is opposed to the direction of the shift, whereas for shifts $1 < \delta < 2$ 
the force acts into the direction of the shift.

At the critical point, i.e., for $\tilde{L}=0$ and $S \to \infty$ in order to keep $\tilde{S}=2$ as in 
Fig.~\ref{fig:ps-tang-SD-S2.0} the patterning becomes irrelevant such that the lateral force attains 
0. With increasing distance $\tilde{L}$ the force reaches a minimum (maximum) for $0<\delta<1$ 
($1<\delta<2$) at a certain $\tilde{L}(\delta) > 0$ and in the limit $\tilde{L} \to \infty$ the force 
vanishes exponentially due to the same arguments which apply to the normal forces. Thus as a function 
of $\tilde{L}$ the lateral force reflects the behavior of the normal force shown in Fig.~\ref{fig:ps-SD_L}. 

\begin{figure}
   \begin{minipage}{0.48\textwidth}
   \begin{center}
      \epsfig{file=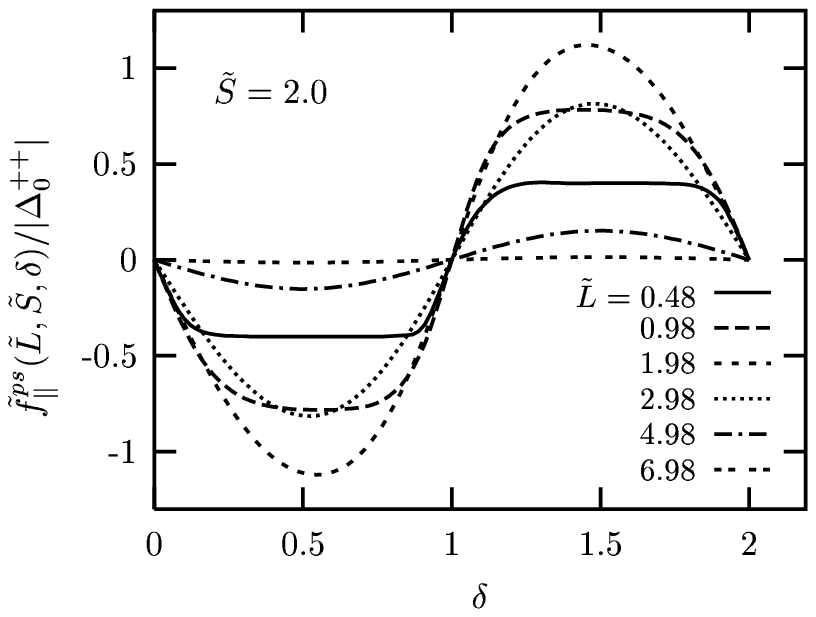,width=\textwidth}
      \caption{\label{fig:ps-tang-SL-S2.0}Scaling function 
               $\tilde{f}^{ps}_\parallel(\tilde{L}, \tilde{S}, \delta)$ of the lateral force normalized 
               by $|\Delta^{++}_0|$ between two identically patterned substrates as a function 
               of the relative shift $\delta$ for different scaled distances $\tilde{L}$ between the 
               substrates and for a scaled stripe width $\tilde{S}=2.0$.}
      \epsfig{file=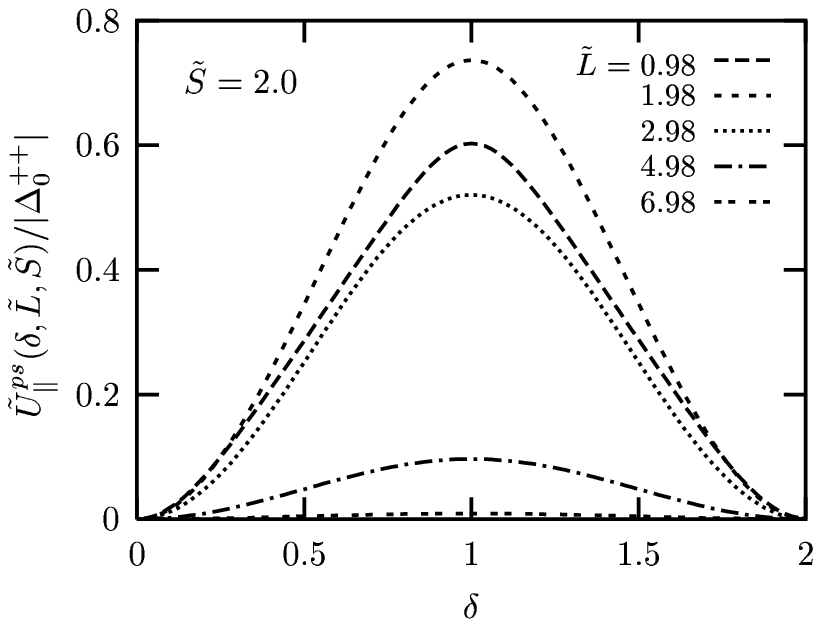,width=\textwidth}
      \caption{\label{fig:energy}The scaling function $\tilde{U}(\delta, \tilde{L}, \tilde{S})$, 
               normalized by $|\Delta^{++}_0|$, of the singular contribution to the cost of free energy 
               to shear the patterned substrates relative to each other by the relative shift $\delta=D/S$. 
               The scaling function has its maximum at $\delta=1$, minima are at $\delta=0$ and $\delta=2$,
               and turning points at $\delta=0.5$ and $\delta=1.5$.}
   \end{center}
   \end{minipage}
\end{figure}

The dependence of the scaling function of the lateral force 
$\tilde{f}^{ps}_\parallel(\tilde{L}, \tilde{S}, \delta)$ on the relative shift $\delta$ is illustrated 
directly in Fig.~\ref{fig:ps-tang-SL-S2.0} for various fixed scaled distances $\tilde{L}$ between the 
substrates and for $\tilde{S}=2$. The force attains a minimum for $\delta \approx 0.5$, i.e., for this 
value the restoring force is strongest. For $\delta \approx 1.5$ the repulsive force attains a maximum. 
For $\tilde{L} \ll \tilde{S}$ the force scaling function exhibit plateaus as function of $\delta$ around 
$\delta =0.5$ and $\delta =1.5$. At $\delta =1$ the force scaling function has an inflection point which 
corresponds to an unstable configuration. \\

The singular free energy cost for shifting the two substrates relative to each other is given by
(see Eq.~(\ref{eq:scaling-force-inhom-ps}))
\begin{equation}
   U^{ps}_\parallel(D, L, S, t) 
   = k_B\,T_c \frac{2NS^2H}{L^d} (d-1) \tilde{U}^{ps}_\parallel(\delta, \tilde{L}, \tilde{S}) \,,
   \label{eq:activation-energy}
\end{equation}
where $N$ is the number of repeat units of width $2S$ and $H$ is the extension of the system in the
translational invariant direction so that $A=2NSH$. The dimensionless scaling function
\begin{equation}
   \tilde{U}^{ps}_\parallel(\delta, \tilde{L}, \tilde{S}) 
   = -\int_0^\delta \dd \delta^\prime \tilde{f}^{ps}_\parallel(\delta^\prime, \tilde{L}, \tilde{S}) \,,
   \label{eq:scal-activation-energy}
\end{equation}
is shown in Fig.~\ref{fig:energy}.

For given lateral features $N$, $S$, and $H$ one has $2NS^2H / L^d \ll 1$ for $L$ sufficiently large 
so that $U^{ps}_\parallel \ll k_B\,T_c$ which implies that the structured substrates can float against 
each other via thermal activation if they are not externally fixed. For $N=50$ repeat units consisting 
of a positive and a negative stripe each of width $S=100$\,nm and for an extension $H=10\,\mu$m the 
above condition is met for $L \gg 2\,\mu$m.

The scaling function $\tilde{U}^{ps}_\parallel(\delta, \tilde{L}, \tilde{S})$ of the free energy exhibits
a maximum $\tilde{U}_{max}$ for $\delta=1$, i.e., $D=S$. This maximum is the activation barrier which 
one has to overcome to shear the patterned system from the $++$~configuration at $\delta=0$ over the 
$+-$~configuration at $\delta=1$ to the $++$~configuration at $\delta=2$. Here the absolute values 
of the free energy do not matter. The turning points at $\delta=0.5$ and $\delta=1.5$ of the scaling 
function of the free energy correspond to the maximal lateral force between the substrates.

\begin{figure}
   \begin{center}
      \epsfig{file=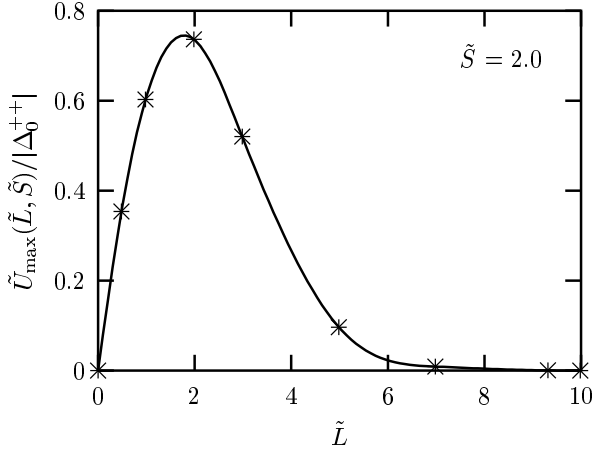,width=0.48\textwidth}
      \caption{\label{fig:energy-max}Scaling function $\tilde{U}_{max}(\tilde{L}, \tilde{S})$, normalized 
               by $|\Delta^{++}_0|$, of the activation barrier as a function of the scaled distance 
               $\tilde{L}$ for a fixed scaled stripe width $\tilde{S}=2.0$.}
   \end{center}
\end{figure}

Figure~\ref{fig:energy-max} shows the universal activation barrier $\tilde{U}_{max}(\tilde{L}, \tilde{S})$ 
with its dependence on the scaled distance $\tilde{L}$ between the substrates for a fixed scaled stripe 
width $\tilde{S}=2$. This scaling function vanishes for $\tilde{L} \to 0$ and $\tilde{L} \to \infty$ and, 
for this choice of $\tilde{S}$, exhibits a maximum for $\tilde{L} \approx 2$.

\subsubsection{Comparison with lateral van der Waals forces}
As described in the Introduction the chemical patterning requires at least one monolayer of the substrate
to be composed of alternating stripes of distinct chemical species which provide the contrast in 
preference for the two types of particles forming the binary liquid mixture exposed to this lateral 
structure. Since these two substrate species interact not only with the fluid particles but also among
each other, their inhomogeneous lateral distribution gives rise to a lateral force even in the absence 
of the fluid between the two confining substrates. This lateral force stems form the direct van der Waals
or dispersion forces between the substrate particles and provides a nonsingular background contribution 
which adds to the singular lateral force discussed above. This nonsingular contribution can be determined
separately by taking the fluid out of the system.

In order to estimate the background force we consider two types of particles $i,j = +,-$ interacting 
pairwise via Lennard-Jones potentials which are characterized by energy and length parameters 
$\epsilon_{ij}$ and $\sigma_{ij}$, respectively. Their actual number density on the stripes are
$\eta_i$. Since $L \gg \sigma_{ij}$, only the attractive part of the Lennard-Jones potentials matters. 
As discussed in Appendix~\ref{sec:append} (see Eq.~(\ref{eq:force-van_der_waals})), the lateral 
van der Waals force (for $d=3$) has the form
\begin{equation}
   \frac{F^{vdW}_\parallel}{2NSH}(D,S,L) 
   = \frac{E}{S^5} \, \tilde{f}^{vdW}_\parallel \left( \frac{L}{S}, \delta \right) \,,
   \label{eq:vdw_scalingbehavior}
\end{equation}
where $E = E_{++} - 2E_{+-} + E_{--}$ and $E_{i,j}=4\epsilon_{ij}\sigma_{ij}^6\eta_i\eta_j$, so that 
$F^{vdW}_\parallel$ vanishes if the chemical contrast disappears, i.e., for $+ =-$; 
$\tilde{f}^{vdW}_\parallel$ is a dimensionless scaling function (see Eq.~(\ref{eq:scaling-van_der_waals})):
\begin{flalign}
   & \tilde{f}^{vdW}_\parallel\left( \frac{L}{S}, \delta \right) 
     = \frac{15\pi}{16} \, \sum_{m=-\infty}^{\infty}
     \label{eq:vdw_scalingfunction} \\
   & \quad \int_0^{1} \dd \zeta_1 \int_{\delta + 2m}^{\delta + 2m+1} \dd \zeta_2
     \frac{\zeta_1-\zeta_2}{ \Big[ \Big( \frac{L}{S} \Big)^2 + (\zeta_1-\zeta_2)^2 \Big]^{\frac{7}{2}} } \,.
   \nonumber 
\end{flalign}

\begin{figure}
   \begin{center}
      \epsfig{file=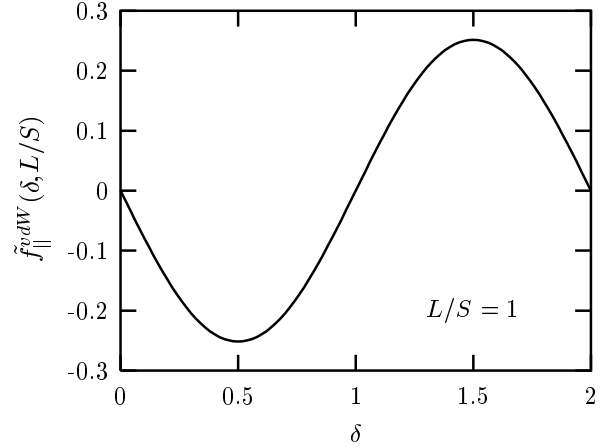,width=0.48\textwidth}
      \caption{\label{fig:vdw-force}Scaling function $\tilde{f}^{vdW}_\parallel$ of the lateral van 
               der Waals force as a function of the relative shift $\delta$ for $L=S$ (see 
               Eqs.~(\ref{eq:vdw_scalingbehavior}) and (\ref{eq:vdw_scalingfunction})).}
   \end{center}
\end{figure}

For $L/S=1$ the scaling function $\tilde{f}^{vdW}_\parallel$ (see Fig.~\ref{fig:vdw-force}) 
attains maximal values of the order of \mbox{$0.3$}. We estimate the prefactor $E/S^5$ as follows.
The interaction energy $\epsilon_{ij}$ is typically in the order of $\epsilon/k_B =300K$ (see Table I 
in Ref.~\onlinecite{Getta1993}), the length $\sigma_{ij}$ is comparable with the atomic diameters of 
the interacting particles, $\sigma_{ij}=0.5$\,nm, and in the case of a \mbox{2-dimensional} closed packing 
the areal number density $\eta_i$ of the particles is $\eta_i=\sqrt{3}/(6 R_i^2)$, with the radius 
$R_i=\sigma_{ii}/2$ of the particles. Assuming stripe widths of $S=100$\,nm one has typical values
\mbox{$E_{ij}/S^5 \simeq 5.5 \cdot 10^{-4}$\,pN/($\mu$m)$^2$}. Thus the lateral van der Waals force 
is of the order of \mbox{$1.7 \cdot 10^{-4}$\,pN/($\mu$m)$^2$}. The actual value is smaller due to 
the differences of $E_{i,j}$ forming $E$.

According to Eq.~(\ref{eq:scaling-force-inhom-ps}) the singular lateral force is given by
\begin{equation}
   \frac{F^{ps}_\parallel(t, L, S, D)}{2NSH}
   = \frac{(d-1)\,k_B\,T_c\,}{L^d} \, 
     \tilde{f}^{ps}_\parallel \left( \tilde{L}, \frac{S}{L}, \delta \right) \,.
 \end{equation}
For a ratio $S/L=1$ the scaling function $\tilde{f}^{ps}_\parallel$ attains maximal values of the 
order of $|\Delta^{++}_0|$ (see Fig.~\ref{fig:ps-tang-SL-S2.0}) with 
$|\Delta^{++}_0|=0.32$.\cite{Krech1997}
For $d=3$, $T_c=300$\,K, and a separation $L=100$\,nm of the substrates, the critical force is of the 
order of $1.3$\,pN/($\mu$m)$^2$. According to this estimate for ratios $S/L \simeq 1$ the direct lateral 
van der Waals force due to the chemical patterning is negligible compared with the critical structural 
force.

\section{\label{sec:summary}Summary}
We have studied the singular contributions to the effective forces acting on chemically inhomogeneous
substrates which confine a binary liquid mixture near a continuous demixing transition. Geometrically 
flat and parallel substrates with periodic chemical patterns of positive stripes (surface field 
$h_1 \to +\infty$) of width $S_+$ and negative stripes ($h_1 \to -\infty$) of width $S_-$ separated 
by a distance $L$ are considered.

Four basic configurations of the two substrates are studied. First, two substrates exhibiting the 
same stripe patterns, i.e., a positive stripe opposite of a positive stripe, which we call 
``$++$~configuration'', are considered (Fig.~\ref{fig:systems}(a), Subsec.~\ref{sec:pp}). Second,
two substrates with opposite stripe patterns, i.e., a positive stripe opposed to a negative one, 
called the ``$+-$~configuration'', are studied (Fig.~\ref{fig:systems}(b), Subsec.~\ref{sec:pm}).
This is followed by investigations of the forces between a structured and a homogeneous substrate, 
called the ``$ph$~configuration'' (Fig.~\ref{fig:systems}(c), Subsec.~\ref{sec:ph}). For two substrates 
with the same but misaligned stripe patterns, called the ``$ps$~configuration'' 
(Fig.~\ref{fig:systems}(d), Subsec.~\ref{sec:ph}), the normal and lateral forces depend on the 
misalignment of the surface structures. 

The universal behavior of the order parameter profiles and of the effective forces acting between the 
substrates is determined by universal scaling functions (Eqs.~(\ref{eq:scaling-force-inhom-con}) - 
(\ref{eq:scaling-force-inhom-ps})). In order to obtain explicit results for these scaling functions 
the order parameter profiles are calculated numerically within mean field theory from which the forces 
between the plates are derived via the stress tensor (Eqs.~(\ref{eq:stress-methode-perp}) - 
(\ref{eq:stress-improv-kompo-perp}), (\ref{eq:f_xz}) - (\ref{eq:stress-improv-kompo-para})). \\

The scaling function of the force between two periodically structured substrates in the 
$++$~configuration depends on the scaled distance $\tilde{L}=L/\xi$ between the substrates, where 
$\xi$ is the bulk correlation length, as well as on the ratio $S_-/S_+$ of the stripe widths 
(Figs.~\ref{fig:pp-SpL1_2-SnSpx} and \ref{fig:pp-SnSp-L}) and the ratio $S_+/L$ 
(Fig.~\ref{fig:pp-SpLx-SnSp2_1}). 
In order to obtain well-defined mean field values for the spatial dimension $d=4$ the scaling functions 
of the force between structured substrates are expressed in units of the universal Casimir amplitude
$\Delta^{++}_0$ characterizing the scaling function of the force between homogeneous substrates with 
parallel surface fields (Fig.~\ref{fig:force-hom_sub} and Eq.~(\ref{eq:casimir-pp})). In the limit 
$S_-/S_+ \to 0$ and $S_-/S_+ \to \infty$ with fixed ratio $S_+/L$ the scaling function of the patterned 
substrates approaches the scaling function of the homogeneous substrates. This limiting case is also 
reached for an increasing ratio $S_+/L$.

At the critical point the force is determined by a universal scaling function of the scaling variables 
$S_-/S_+$ and $S_+/L$, which reduces to the Casimir amplitude $\Delta^{++}_0$ in the limits 
$S_-/S_+ \to 0$ and $S_-/S_+ \to \infty$ (Fig.~\ref{fig:casimir-pp-SnSp}). The limit $L \to \infty$ 
of this scaling function can be interpreted as a generalized Casimir amplitude $\Delta^{++}(S_-/S_+)$. 
Thus the chemical patterning of the substrates allows one to tune the universal Casimir amplitude within 
a finite range of negative values (Fig.~\ref{fig:casimir-pp-SnSp}). \\

The force between two periodically structured substrates in the $+-$~configuration is governed by
a scaling function which depends on the same scaling variables $\tilde{L}$, $S_-/S_+$ 
(Figs.~\ref{fig:pm-SpL1_2-SnSpx} and \ref{fig:pm-SnSp-L}), and $S_+/L$ (Fig.~\ref{fig:pm-SpLx-SnSp2_1}).
Like for the $++$~configuration it is reduced compared to the scaling function of the force between 
two homogeneous substrates with antiparallel surface fields. Analogously to the $++$~configuration, 
the scaling function of the $+-$~configuration attains the scaling function of the homogeneous substrates 
in the limits $S_-/S_+ \to 0$ and $S_-/S_+ \to \infty$ with a fixed ratio $S_+/L$ and in the limit 
$S_+/L\to \infty$ for a fixed ratio $S_-/S_+$. \\

As in the case of the $++$~configuration, a generalized Casimir amplitude $\Delta^{+-}(S_-/S_+)$ 
can be defined also for the $+-$~configuration. This follows from the universal scaling function 
describing the force at $T_c$ which depends on the scaling variables $S_-/S_+$ and $S_+/L$. In the 
limits $S_-/S_+ \to 0$ and $S_-/S_+ \to \infty$ it reduces to the Casimir amplitude $\Delta^{+-}_0 > 0$ 
(Eq.~(\ref{eq:casimir-pm})) characterizing the force between homogeneous substrates with antiparallel
surface fields (Fig.~\ref{fig:casimir-pm-SnSp}). Here the chemical patterning allows to tune the 
universal Casimir amplitude within a finite range of positive values (Fig.~\ref{fig:casimir-pm-SnSp}).
The ratio of the generalized Casimir amplitudes for the $+-$ and the $++$~configuration  as function
of $S_-/S_+$ is shown in Fig.~\ref{fig:compare-amplitudes}. \\

The $ph$~configuration of a critical fluid confined between a periodically patterned and a 
homogeneous substrate with positive surface field interpolates, as a function of the ratio $S_-/S_+$ 
of the stripe widths, between the case of two homogeneous substrates with parallel surface fields 
($S_-/S_+ \to 0$) and the case of two homogeneous substrates with antiparallel surface fields 
($S_-/S_+ \to \infty$) (Fig.~\ref{fig:ph-SpL1_2-SnSpx}, Figs.~\ref{fig:ph-SpLx-SnSp1_2} - 
\ref{fig:ph-SpLx-SnSp2_1}). Accordingly the corresponding generalized Casimir amplitude 
$\Delta^{ph}(S_-/S_+)$ interpolates between the Casimir amplitude $\Delta^{++}_0 < 0$ at $S_-/S_+=0$ 
and the Casimir amplitude $\Delta^{+-}_0 > 0$ at $S_-/S_+ \to \infty$. Figure~\ref{fig:casimir-ph-SnSp} 
shows universal scaling functions at $T_c$ for non-vanishing ratios $S_+/L$, which converge to the 
generalized Casimir amplitude $\Delta^{ph}(S_-/S_+)$ in the limit $S_+/L \to 0$. In this case the chemical
patterning allows one to tune even the sign of the Casimir amplitude and to drive the leading critical
Casimir force to zero. This opens the possibility to observe the leading correction term which is
typically masked by the dominant term. \\

The absolute value of the scaling function of the force between substrates in the $++$~configuration
is smaller than the corresponding one for the $+-$~configuration (Fig.~\ref{fig:vgl-SpLx-SnSpy}).
In the case of very narrow stripes, i.e., for $S_+/L \ll 1$ 
(Figs.~\ref{fig:vgl-SpLx-SnSpy}(b) and (d)) the absolute value of the scaling function of the 
$ph$~configuration is significantly larger than the one for the $++$~configuration and the one for the 
$+-$~configuration. The reason for this is that the surface fields of the structured substrates 
effectively tend to zero for decreasing stripe widths -- mimicking Dirichlet boundary conditions 
-- whereas the surface field of the homogeneous substrate of the $ph$~configuration remains finite. \\

If the structures of the patterned substrates are not in phase but shifted relative to each other, 
an additional lateral force emerges which acts as to restore a configuration in which the structures 
are again in phase. In this $ps$~configuration the scaling functions of the force depend additionally 
on the relative shift $\delta=D/S$ of the stripes of one substrate with respect to the stripes of the 
other substrate (Fig.~\ref{fig:systems}(d)). In order to limit the number of the relevant length 
scales here we have chosen $S_-=S_+=S$. The scaling function of the normal force as a function of the 
relative shift $\delta$ transfers the scaling function for the $++$~configuration for $\delta = 0$ into 
the scaling function for the $+-$~configuration for $\delta = 1$ (Figs.~\ref{fig:ps-SD_L} and 
\ref{fig:ps-SL_D}). 

The importance of the relative shift $\delta$ decreases with increasing scaled distance $\tilde{L}$ 
between the substrates. The influence of the scaled stripe width $\tilde{S}=S/\xi$ is visible through 
the dependence of the scaling function on the ratio $L/S$ (Fig.~\ref{fig:ps-S1S2D_L-scaled}) with the 
conclusion that the scaling function does not reduce to an effective scaling function of only the two 
variables $L/S$ and $\delta$. 

The generalized Casimir amplitude $\Delta^{ps}(S/L,\delta)$, characterizing the scaling function of 
the normal force at the critical temperature (Fig.~\ref{fig:casimir-ps-SL1_x}), vanishes in the limit 
$S/L \to 0$. For positive and negative stripes of the same width, in this limit the surface field 
effectively tends to zero corresponding to Dirichlet (ordinary) boundary conditions for which the force 
vanishes within mean field theory. \\

Figures~\ref{fig:casimir-pp-SnSp}, \ref{fig:casimir-pm-SnSp}, \ref{fig:casimir-ph-SnSp}, and 
\ref{fig:casimir-ps-SL1_x} demonstrate the tunability of the Casimir amplitudes via the chemical 
patterning of the substrates.
The fact that within mean field theory the Casimir amplitudes $\Delta^{++}(S_-/S_+, S_+/L)$, 
$\Delta^{+-}(S_-/S_+, S_+/L)$, and $\Delta^{ps}(\delta, S/L)$ vanish for $S_+/S_-=1 $ in the limit 
$S_+/L \to 0$ and $S/L \to 0$, respectively, resembling Dirichlet boundary conditions, is 
remarkable. In general in classical fluid systems the substrate potential breaks the symmetry of the 
order parameter such that these systems are unable to exhibit Dirichlet boundary conditions. Only in 
the case of quantum fluids, i.e., in critical superfluid He$^4$ films, Dirichlet boundary conditions 
show up.\cite{Krech1992b, Garcia1999} The results shown in Figs.~\ref{fig:casimir-pp-SnSp}, 
\ref{fig:casimir-pm-SnSp}, \ref{fig:casimir-ph-SnSp}, and \ref{fig:casimir-ps-SL1_x} visualize that 
for suitably chosen chemical patterns Dirichlet boundary conditions effectively emerge even for classical 
fluids. \\

The scaling function of the lateral force vanishes for relative shifts $\delta=0$, $\delta=1$, and 
$\delta=2$ (Figs.~\ref{fig:ps-tang-SD-S2.0} and \ref{fig:ps-tang-SL-S2.0}). For shifts $0 < \delta < 1$
the scaling function of the lateral force is negative, i.e., restoring. For $1 < \delta < 2$ it is 
positive, i.e., the force acts into the direction of the shift. For $\delta=1$ the scaling function 
exhibits an inflection point, which means that the corresponding $+-$~configuration is an unstable 
configuration.

The free energy required to shift the patterned substrates relative to each other 
(Eq.~(\ref{eq:activation-energy})) is given by a scaling function (Eq.~(\ref{eq:scal-activation-energy}) 
and Fig.~\ref{fig:energy}) which shows a maximum at $D \simeq S$ and exhibits inflection points at
$D/S=0.5$ and $D/S=1.5$ corresponding to the occurrence of maximal forces. The maximum of this free 
energy, i.e., the activation energy needed to move a system from a $++$~configuration to the adjacent
$+-$~configuration via shifting the substrates by $D=S$, depends on the distance between the 
substrates and vanishes for increasing distances (Fig.~\ref{fig:energy-max}).

The background contribution from direct van der Waals forces (Fig.~\ref{fig:vdw-force}), generated 
by the chemical patterning within a monolayer covering the substrates, is negligible for ratios 
\mbox{$S/L \simeq 1$} compared with the critical lateral force.

\appendix
\section{\label{sec:append}Lateral van der Waals force}
The lateral van der Waals force $F^{vdW}_\parallel = - \frac{\partial U^{vdW}_\parallel}{\partial D}$
is generated by the direct interaction between the two substrate species forming the chemical pattern.
Here we assume that this chemical pattern is confined to a monolayer and that the corresponding pair
potentials are given by Lennard-Jones potentials:
\begin{equation}
   U_{ij}= 4 \epsilon_{ij} \left[ \left(\frac{\sigma_{ij}}{r}\right)^{12}
                                 -\left(\frac{\sigma_{ij}}{r}\right)^6 \right] \,,
   \quad i,j \in \{ +,- \} \,.
\end{equation}
Since $L \gg \sigma_{ij}$, for the lateral van der Waals force only the attractive part of the 
Lennard-Jones potentials is important. Assuming pairwise additivity the leading contribution to 
$U^{vdW}_\parallel$ is given by:\cite{Karimi2005}
\begin{flalign}
   & U^{vdW}_\parallel = - \int_{S_1} \dd x_1 \dd y_1 \int_{S_2} \dd x_2 \dd y_2 \nonumber \\
   & \hspace{20mm} \frac{E(x_1, x_2, D)}{[L^2+(x_1-x_2)^2+(y_1-y_2)^2]^3} \,,
   \label{eq:uvdw}
\end{flalign}
with $x_1$ and $x_2$ denoting the direction of the inhomogeneity on the substrate $S_1$ and $S_2$, 
respectively, and $y_1$ and $y_2$ the directions perpendicular to it (see Fig.~\ref{fig:systems}). 
Depending on the lateral position, $E(x_1, x_2, D)$ equals $E_{++}$, $E_{+-}$, or $E_{--}$ (see, 
c.f., Eqs.~(\ref{eq:epp_epm}) and (\ref{eq:epm_emm})).
The contributions $E_{ij}$ are given by $E_{ij}=4\epsilon_{ij}\sigma_{ij}^6\eta_i\eta_j$, where 
$\eta_i$ denote the areal number densities of the particles. \\

In order to carry out the quadruple integral in Eq.~(\ref{eq:uvdw}) we first consider the interaction 
$U^{vdW}_{\parallel,S_+}$ of a positive stripe on the substrate $S_1$ of width $S$ and extension $H$ 
in the $y_1$~direction with the whole substrate $S_2$ with infinite extensions in the $x_2$ and 
$y_2$~directions:
\begin{flalign}
   & U^{vdW}_{\parallel,S_+}
     = - \int_0^S \dd x_1 \int_0^H \dd y_1 
         \int_{-\infty}^{\infty} \dd x_2 \int_{-\infty}^{\infty} \dd y_2 \, \nonumber \\
   & \hspace{20mm} \frac{E(x_1, x_2, D)}{[L^2+(x_1-x_2)^2+(y_1-y_2)^2]^3} \,.
   \label{eq:int_y2}
\end{flalign}
With
\begin{eqnarray}
   \int_{-\infty}^{+\infty} \dd x \frac{1}{(a^2+x^2)^3}
   &=& \frac{3\pi}{8a^5}
\end{eqnarray}
Eq.~(\ref{eq:int_y2}) gives
\begin{eqnarray}
   U^{vdW}_{\parallel,S_+}
   = - \frac{3\pi}{8} \, H \int_0^S \dd x_1 \int_{-\infty}^{\infty} \dd x_2 \,
            \frac{E(x_1, x_2, D)}{[L^2+(x_1-x_2)^2]^{\frac{5}{2}}} \,. \nonumber \\
   \label{eq:int_x1x2}
\end{eqnarray}
The integration over the $x_2$ direction can be split up into integrations over the stripes of the
substrates $S_2$, for which the strength $E(x_1,x_2,D)$ of the interaction is constant:
\begin{flalign}
   & x_1 \in [0 \; , \; S] \,, \; x_2 \in [D + 2mS \; , \; D + (2m+1)S]: \nonumber \\[1mm]
   & \quad E(x_1,x_2,D) =  E_{++} \nonumber \\[3mm]
   & x_1 \in [0 \; , \; S] \,, \; x_2 \in [D + (2m+1)S \; , \; D + (2m+2)S]: \nonumber \\
   & \quad E(x_1,x_2,D) =   E_{+-} \,,
   \label{eq:epp_epm}
\end{flalign}
where $m \in \mathbb{Z}$.
Thus one has for $U^{vdW}_{\parallel,S_+}$ (Eq.~(\ref{eq:int_x1x2})):
\begin{flalign}
   & U^{vdW}_{\parallel,S_+} 
     = - \frac{3\pi}{8} \, H \sum_{m=-\infty}^{\infty}  \nonumber \\
   & \left( \int_0^S \dd x_1 \int_{D + 2mS}^{D + (2m+1)S} \dd x_2 \,
            \frac{E_{++}}{[L^2+(x_1-x_2)^2]^{\frac{5}{2}}} 
     \right. \nonumber \\
   & \left. + \int_0^S \dd x_1 \int_{D + (2m+1)S}^{D + (2m+2)S} \dd x_2 \,
            \frac{E_{+-}}{[L^2+(x_1-x_2)^2]^{\frac{5}{2}}} 
     \right) \,.
   \label{eq:u_vdw_s_p}
\end{flalign}

Now we consider the interaction energy $U^{vdW}_{\parallel,S_-}$ of a negative stripe on the substrate 
$S_1$ of width $S$ and extension $H$ in the $y_1$~direction with the whole substrate $S_2$ with infinite 
extensions in $x_2$ and $y_2$~direction. The strength of the interaction is given by
\begin{flalign}
   & x_1 \in [S \; , \; 2S] \,, \; x_2 \in [D + 2mS \; , \; D + (2m+1)S]: \nonumber \\[1mm]
   & \quad E(x_1,x_2,D) = E_{+-} \nonumber \\[3mm]
   & x_1 \in [S \; , \; 2S] \,, \; x_2 \in [D + (2m+1)S\; , \; D + (2m+2)S]: \nonumber \\[1mm]
   & \quad E(x_1,x_2,D) = E_{--} \,,
   \label{eq:epm_emm}
\end{flalign}
such that $U^{vdW}_{\parallel,S_-}$ is described by:
\begin{flalign}
   & U^{vdW}_{\parallel,S_-}
     = - \frac{3\pi}{8} \, H \sum_{m=-\infty}^{\infty}  \nonumber \\
   & \left( \int_{S}^{2S} \dd x_1 \int_{D + 2mS}^{D + (2m+1)S} \dd x_2 \,
            \frac{E_{+-}}{[L^2+(x_1-x_2)^2]^{\frac{5}{2}}} 
         \right. \nonumber \\
   & \left. + \int_{S}^{2S} \dd x_1 \int_{D + (2m+1)S}^{D + (2m+2)S} \dd x_2 \,
            \frac{E_{--}}{[L^2+(x_1-x_2)^2]^{\frac{5}{2}}}
         \right) \,.
   \label{eq:u_vdw_s_n}
\end{flalign}

The sum of the interaction energies $U^{vdW}_{\parallel,S_+}$ (Eq.~(\ref{eq:u_vdw_s_p})) and 
$U^{vdW}_{\parallel,S_-}$ (Eq.~(\ref{eq:u_vdw_s_n})) is
\begin{flalign}
    & U^{vdW}_{\parallel,S_+} + U^{vdW}_{\parallel,S_-}
      = - \frac{3\pi}{8} \, H \sum_{m=-\infty}^{\infty} \nonumber \\
    & \quad 
      \left( \int_{D + 2mS}^{D + (2m+1)S} \dd x_2 \left( 
             \int_0^S \dd x_1 \frac{E_{++}}{[L^2+(x_1-x_2)^2]^{\frac{5}{2}}} 
      \right. 
      \right. \nonumber \\
    & \left. \hspace{33mm} + \int_S^{2S} \dd x_1 \frac{E_{+-}}{[L^2+(x_1-x_2)^2]^{\frac{5}{2}}}
      \right) \hspace{1cm} \nonumber \\
    & \quad 
      + \int_{D + (2m+1)S}^{D + (2m+2)S} \dd x_2 \left(
               \int_0^S \dd x_1 \frac{E_{+-}}{[L^2+(x_1-x_2)^2]^{\frac{5}{2}}}
      \right. \nonumber \\
    & \left. 
      \left. \hspace{33mm} + \int_S^{2S} \dd x_1 \frac{E_{--}}{[L^2+(x_1-x_2)^2]^{\frac{5}{2}}}
      \right) 
      \right) \nonumber \\
    & = - \frac{3\pi}{8} \, H \sum_{m=-\infty}^{\infty} \nonumber \\
    & \quad \left( \int_{D + 2mS}^{D + (2m+1)S} \dd x_2 \int_0^S \dd x_1 \,
                 \frac{E_{++} - E_{+-}}{[L^2+(x_1-x_2)^2]^{\frac{5}{2}}} 
              \right.  \nonumber \\[1mm]
    & \quad + \int_{D + 2mS}^{D + (2m+1)S} \dd x_2 \int_0^{2S} \dd x_1 \,
                 \frac{E_{+-}}{[L^2+(x_1-x_2)^2]^{\frac{5}{2}}} 
              \nonumber \\[1mm]
    & \quad + \int_{D + (2m+1)S}^{D + (2m+2)S} \dd x_2 \int_S^{2S} \dd x_1 \,
                 \frac{E_{--} - E_{+-}}{[L^2+(x_1-x_2)^2]^{\frac{5}{2}}} 
              \nonumber \\[1mm]
    & \quad \left. + \int_{D + (2m+1)S}^{D + (2m+2)S} \dd x_2 \int_0^{2S} \dd x_1 \,
                 \frac{E_{+-}}{[L^2+(x_1-x_2)^2]^{\frac{5}{2}}} 
              \right) \,.
\end{flalign}
The transformation $X_1=x_1-S_+$ and $X_2=x_2-S_+$ in the third term renders:
\begin{flalign}
   & U^{vdW}_{\parallel,S_+} + U^{vdW}_{\parallel,S_-} 
     = - \frac{3\pi}{8} \, H \sum_{m=-\infty}^{\infty} \nonumber \\
   & \left( \int_{D + 2mS}^{D + (2m+1)S} \dd x_2 \int_0^S \dd x_1 \,
                 \frac{E_{++} - 2E_{+-} + E_{--}}{[L^2+(x_1-x_2)^2]^{\frac{5}{2}}} 
              \right. \nonumber \\[1mm]
   & \quad \left. + \int_{D + 2mS}^{D + (2m+2)S} \dd x_2 \int_0^{2S} \dd x_1 \,
                 \frac{E_{+-}}{[L^2+(x_1-x_2)^2]^{\frac{5}{2}}} 
              \right) \,. 
\end{flalign}
Scaling the lengths with the stripe width $S$,
\begin{equation}
   \zeta_1 = \frac{x_1}{S} \,, \qquad \zeta_2 = \frac{x_2}{S} \,, \qquad \delta= \frac{D}{S} \,,
\end{equation}
leads to
\begin{flalign}
   & U^{vdW}_{\parallel,S_+} + U^{vdW}_{\parallel,S_-} 
     = - \frac{3\pi}{8} \, \frac{H}{S^3} \sum_{m=-\infty}^{\infty} \nonumber \\
   & \left( \int_{\delta + 2m}^{\delta + (2m+1)} \dd \zeta_2 \int_0^1 \dd \zeta_1 \, 
             \frac{E_{++} - 2E_{+-} + E_{--}}
                  { \Big[ \Big( \frac{L}{S} \Big)^2 + (\zeta_1-\zeta_2)^2 \Big]^{\frac{5}{2}} }
     \right. \nonumber \\[1mm]
   & \quad \left. + \int_{\delta + 2m}^{\delta + (2m+2)} \dd \zeta_2 \int_0^{2} \dd \zeta_1 \,
                  \frac{E_{+-}}
                       { \Big[ \Big( \frac{L}{S} \Big)^2 + (\zeta_1-\zeta_2)^2 \Big]^{\frac{5}{2}} }
          \right) \,. 
\end{flalign}
The sum over the second integral reduces to:
\begin{flalign}
   & \sum_{m=-\infty}^{\infty} 
     \int_{\delta + 2m}^{\delta + (2m+2)} \dd \zeta_2 \int_0^{2} \dd \zeta_1 \,
        \frac{E_{+-}}{ \Big[ \Big( \frac{L}{S} \Big)^2 + (\zeta_1-\zeta_2)^2 \Big]^{\frac{5}{2}} }
     \nonumber \\
   & = \lim_{m \to \infty}
       \int_{\delta - 2m}^{\delta + (2m+2)} \dd \zeta_2 \int_0^{2} \dd \zeta_1 \,
           \frac{E_{+-}}{ \Big[ \Big( \frac{L}{S} \Big)^2 + (\zeta_1-\zeta_2)^2 \Big]^{\frac{5}{2}} } \,.
     \nonumber \\
\end{flalign}

\begin{widetext}
The interaction energy $U^{vdW}_\parallel/N$ of a positive and a negative stripe on one substrate, where 
$N$ is the number of repeat units, with the structure on the other substrate per area $2SH$ reads
\begin{eqnarray}
    \frac{U^{vdW}_\parallel}{2NSH}
    &=& - \frac{3\pi}{16} \, \frac{E}{S^4} \sum_{m=-\infty}^{\infty}
         \int_0^{1} \! \dd \zeta_1 \int_{\delta + 2m}^{\delta + 2m+1} \! \dd \zeta_2 \,
           \frac{1}{ \Big[ \Big( \frac{L}{S} \Big)^2 + (\zeta_1-\zeta_2)^2 \Big]^{\frac{5}{2}} } 
         \nonumber \\
    & & - \frac{3\pi}{16} \, \frac{E_{+-}}{S^4} \lim_{m \to \infty} 
           \int_0^{2} \! \dd \zeta_1 \int_{\delta - 2m}^{\delta + (2m+2)} \! \dd \zeta_2 \,
               \frac{1}{ \Big[ \Big( \frac{L}{S} \Big)^2 + (\zeta_1-\zeta_2)^2 \Big]^{\frac{5}{2}} } \,,
\end{eqnarray}
where $E = E_{++} - 2E_{+-} + E_{--}$. \\
The lateral van der Waals force is determined by the derivative of $U^{vdW}_\parallel$ with respect to
the shift $D$:
\begin{eqnarray}
    \frac{F^{vdW}_\parallel}{2NSH}
   = - \frac{1}{2NSH} \frac{\partial U^{vdW}_\parallel}{\partial D} 
   &=& \frac{3\pi}{16} \, \frac{E}{S^5}
       \sum_{m=-\infty}^{\infty} \frac{\partial}{\partial \delta} 
       \int_0^{1} \dd \zeta_1 \int_{\delta + 2m}^{\delta + 2m+1} \dd \zeta_2 \,
           \frac{1}{ \Big[ \Big( \frac{L}{S} \Big)^2 + (\zeta_1-\zeta_2)^2 \Big]^{\frac{5}{2}} } 
       \nonumber \\[2mm]
   & & + \frac{3\pi}{16} \, \frac{E_{+-}}{S^5} \lim_{m \to \infty} \frac{\partial}{\partial \delta} \Bigg\{
           \int_0^{2} \dd \zeta_1 \int_{\delta - 2m}^{\delta + (2m+2)} \dd \zeta_2 \,
               \frac{1}{ \Big[ \Big( \frac{L}{S} \Big)^2 + (\zeta_1-\zeta_2)^2 \Big]^{\frac{5}{2}} }
        \Bigg\} \nonumber \\
   &=& \frac{15\pi}{16} \, \frac{E}{S^5}
       \sum_{m=-\infty}^{\infty}
       \int_0^{1} \dd \zeta_1 \int_{\delta + 2m}^{\delta + 2m+1} \dd \zeta_2 \,
          \frac{\zeta_1-\zeta_2}{ \Big[ \Big( \frac{L}{S} \Big)^2 + (\zeta_1-\zeta_2)^2 \Big]^{\frac{7}{2}} }
       \nonumber \\[1mm]
   && + \frac{15\pi}{16} \, \frac{E_{+-}}{S^5} \lim_{m \to \infty}
       \int_0^{2} \dd \zeta_1 \int_{\delta - 2m}^{\delta + (2m+2)} \dd \zeta_2 \,
           \frac{\zeta_1-\zeta_2}
                { \Big[ \Big( \frac{L}{S} \Big)^2 + (\zeta_1-\zeta_2)^2 \Big]^{\frac{7}{2}} } \,.
        \label{eq:herleit_vdw2}
\end{eqnarray}
The last term in Eq.~(\ref{eq:herleit_vdw2}) vanishes:
\begin{equation}
    \lim_{m \to \infty}
          \int_0^{2} \dd \zeta_1 \int_{\delta - 2m}^{\delta + (2m+2)} \dd \zeta_2 \,
          \frac{\zeta_1-\zeta_2}{ \Big[ \Big( \frac{L}{S} \Big)^2 + (\zeta_1-\zeta_2)^2 \Big]^{\frac{7}{2}} }
    = 0 \,.
\end{equation}
Therefore the lateral van der Waals force $F^{vdW}_\parallel$ per area $2NSH$ between two structured 
monolayers is given by
\begin{equation}
   \frac{F^{vdW}_\parallel}{2NSH} = \frac{E}{S^5} \, \tilde{f}^{vdW}_\parallel
   \label{eq:force-van_der_waals}
\end{equation}
with the scaling function
\begin{equation}
   \tilde{f}^{vdW}_\parallel
   = \frac{15\pi}{16} \, \sum_{m=-\infty}^{\infty}
     \int_0^{1} \dd \zeta_1 \int_{\delta + 2m}^{\delta + 2m+1} \dd \zeta_2 \,
     \frac{\zeta_1-\zeta_2}{ \Big[ \Big( \frac{L}{S} \Big)^2 + (\zeta_1-\zeta_2)^2 \Big]^{\frac{7}{2}} } \,.
   \label{eq:scaling-van_der_waals}
\end{equation}
\end{widetext}


\end{document}